\documentclass[review,a4paper,3p,12pt]{elsarticle}
\usepackage{amsmath,amssymb,amsthm}
\usepackage{mathtools}
\usepackage{bm}
\usepackage{graphicx}
\usepackage{hyperref}
\usepackage{booktabs}
\usepackage{array}
\usepackage{enumitem}
\usepackage{xcolor}
\graphicspath{{./}}
\newcommand{\E}{\mathbb{E}}
\newcommand{\Var}{\mathrm{Var}}
\newcommand{\Cov}{\mathrm{Cov}}
\newcommand{\eff}{\mathrm{eff}}
\newcommand{\const}{\mathrm{const}}

\newcommand{\prob}{\mathbb{P}}
\newcommand{\diff}{\mathrm{d}}
\newcommand{\arctanh}{\mathrm{arctanh}}
\DeclareMathOperator{\Tr}{Tr}
\newcolumntype{L}[1]{>{\raggedright\arraybackslash}p{#1}}

\begin{document}

\begin{frontmatter}

\title{Exact log-odds representation and mean-field criticality
of a growing social group model}

\author[hnu]{Xingfu Ke}
\author[hnu]{Fanyuan Meng\corref{corresponding}}
\cortext[corresponding]{Corresponding author: fanyuan.meng@hotmail.com}
\address[hnu]{Alibaba Research Center for Complexity Sciences,
Hangzhou Normal University, Hangzhou, Zhejiang 311121, China}

\begin{abstract}
We present an exact analytical reformulation of a growing social group
model---a Hamiltonian-free nonequilibrium process in which a group grows
by noisy, consensus-driven admission.
Cast as a gradient flow on logarithmic time, the fixed-point structure
collapses to a single self-consistent equation:
$\arctanh(\phi^*) = m \cdot \arctanh(\alpha\phi^*)$,
where $\phi$ is the polarization, $\alpha=1-2\eta$ the evaluation
reliability, and $m$ the number of evaluators.
The equation has a direct log-odds interpretation: each verdict
contributes log-likelihood ratio $2\arctanh(\alpha\phi)$; unanimity
accumulates $m$ independent evidence pieces. The dynamics thus
constitutes an exact mean-field theory of self-consistent inference,
ordering when the collective gain $m\alpha$ overcomes the dilution
of growth.
We develop a systematic three-layer framework:
core theory (Landau-like effective potential, comparison with the
mean-field Ising model, and features without equilibrium counterpart),
mathematical foundations (criticality from correlated verdicts,
P\'{o}lya-urn martingale convergence, and an RG-like flow with group size
as scale), and complementary perspectives on irreversibility and
information geometry.
A frozen-$N$ Freidlin--Wentzell quasipotential yields Kramers-type
escape estimates for metastable states, while Monte Carlo simulations
collapse onto a parameter-free deterministic master curve on logarithmic
time.
Systematic comparison with the mean-field Ising model reveals shared
critical exponents but a nested arctanh structure unique to growth.
These results provide a detailed analytical characterization of a minimal
model of growth-driven collective behavior and map which elements of the
equilibrium critical toolbox---suitably reinterpreted---survive without a
Hamiltonian.
\end{abstract}


\end{frontmatter}
	
	\section{Introduction}
	\label{sec:intro}
	
	Critical phenomena---the emergence of singular thermodynamic behavior at a
	critical point---form the conceptual backbone of modern statistical physics
	\cite{Stanley1971,Ma1976,Goldenfeld1992,Nishimori2010}.
	The standard theoretical framework rests on three pillars:
	(i)~a microscopic Hamiltonian $H(\{\sigma\})$ encoding interactions;
	(ii)~the Boltzmann--Gibbs distribution $P(\{\sigma\}) \propto e^{-\beta H}$
	defining thermal equilibrium;
	and (iii)~the thermodynamic limit $N\to\infty$ revealing singularities in
	the free energy $F = -k_B T \ln Z$.
	This framework, exemplified by the Ising model
	\cite{Ising1925,Onsager1944,Yang1952}, is remarkably general and has been
	applied far beyond its original domain of magnetism---to opinion
	dynamics \cite{Galam2002,Castellano2009,Medo2021}, cultural diffusion
	\cite{Axelrod1997}, ecological phase transitions \cite{Sole2011}, and
	collective animal behavior \cite{Bialek2012}.
	
	Yet many complex systems of current interest are \emph{intrinsically
		nonequilibrium}: they evolve by growth, replication, or sequential addition
	of constituents, without ever visiting a Boltzmann distribution.
	Examples include the assembly of ecological communities
	\cite{Hubbell2001,Azaele2016}, the growth of social networks and online groups
	\cite{Backstrom2006,Leskovec2008}, the accumulation of citations in science
	\cite{Price1976,Barabasi1999}, the diffusion of innovations
	\cite{Rogers2003}, the clonal expansion of cell populations
	\cite{Nowell1976}, and the expansion of firms and cities
	\cite{Gabaix1999,Bettencourt2007}.
	In such systems, there is no Hamiltonian, no temperature, and no a priori
	reason for a free-energy-like functional to exist.
	Can critical phenomena nonetheless emerge?
	
	The question is not merely academic.
	Empirically, growing groups---political parties, research teams, online
	communities---exhibit a tension between cohesion and size
	\cite{Delhey2007,Wheelan2009}.
	Small groups tend to be cohesive; large groups tend to fragment.
	The mechanism by which growth erodes cohesion, and the conditions under
	which cohesion can persist, are central to understanding organizational
	stability \cite{Carron2002,Bruhn2009}.
	Recently, Fenoaltea, Meng, Liu, and Medo (FMLM)
	\cite{Fenoaltea2023} proposed a minimalist model in which a group grows by
	sequentially admitting or rejecting candidates of two types through a noisy,
	consensus-based evaluation process.
	The model exhibits a rich phase diagram: when candidates are evaluated by
	a single group member ($m=1$), any nonzero evaluation noise drives cohesion
	to the random-group level as $N\to\infty$; when consensus of $m\ge 2$ members
	is required, a critical noise threshold $\eta_c = 1/2 - 1/(2m)$ separates a
	cohesive phase from a disordered one; and when unfit candidates are more
	probable than fit ones ($f<1/2$), a discontinuous, history-dependent
	transition appears.
	
\begin{figure*}[t]
		\centering
		\includegraphics[width=0.82\textwidth]{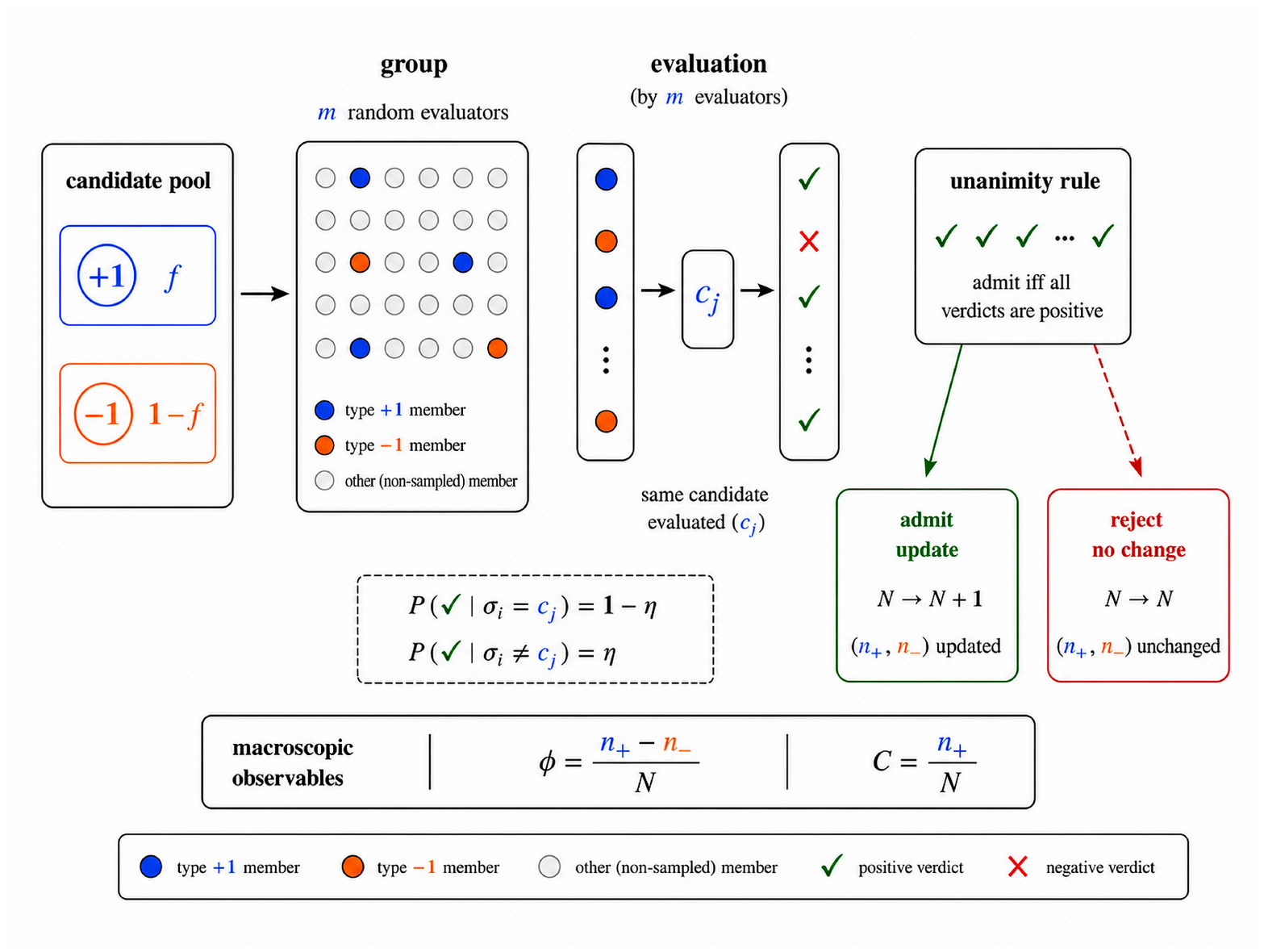}
		\caption{Schematic of the growing-group model.
			A candidate drawn from the pool (type $+1$ with probability $f$,
			type $-1$ with probability $1-f$) is evaluated by $m$ members
			chosen uniformly at random from the current group.
			All $m$ evaluators judge the same candidate $c_j$: each
			independently returns a positive verdict with probability $1-\eta$
			if its own type matches the candidate's type, and with probability
			$\eta$ otherwise.
			Under the unanimity rule the candidate is admitted if and only if
			all $m$ verdicts are positive, whereupon $N\to N+1$ and the
			composition $(n_+,n_-)$ is updated; otherwise the group is left
			unchanged.
			The macroscopic observables are the polarization
			$\phi=(n_+-n_-)/N$ and the cohesion $C=n_+/N$.}
		\label{fig:schematic}
	\end{figure*}

	In this paper, we revisit the FMLM model from the perspective of nonequilibrium
	statistical mechanics.
	Our contribution is not a new model, but a systematic analytical framework
	for understanding it.
	To delineate the boundary with Ref.~\cite{Fenoaltea2023} explicitly: the
	existence of the transition, the critical noise $\eta_c=1/2-1/(2m)$, the
	closed-form $m=2$ solution, and the discontinuous transition for $f<1/2$
	were all established there.
	What is new here is
	(i)~the arctanh representation of the fixed-point equation, its
	identification as the exact log-odds (Bayesian-inference) variable of the
	evaluation process (Sec.~\ref{sec:bayes}), and the explicit Landau
	coefficients that follow from it;
	(ii)~the identification of the candidate bias $h=\frac12\ln[f/(1-f)]$ as an
	exact symmetry-breaking field, and the resulting demonstration that
	$(f,\eta)=(1/2,\eta_c)$ is an ordinary critical point rather than a
	tricritical point, with the testable critical isotherm
	$|\phi^*|\propto|f-1/2|^{1/3}$;
	(iii)~the logarithmic-time relaxation-scaling form, its Monte Carlo
	verification, and a fluctuation-level test of the stochastic sector;
	(iv)~the martingale convergence statement for the stochastic process; and
	(v)~the quasipotential, escape-rate, and nonlinear-susceptibility
	predictions.
	Table~\ref{tab:provenance} summarizes this division of provenance.
	We also place the model in the context of phase transitions in
	Hamiltonian-free growth processes, which have a substantial history:
	nonlinear P\'{o}lya urns \cite{Hill1980,Pemantle2007}, condensation in
	growing networks \cite{Bianconi2001,Krapivsky2000}, and absorbing-state
	transitions in driven systems \cite{Hinrichsen2000}.
	Relative to this literature, the present model is distinguished by being
	exactly solvable at the mean-field level while exhibiting both continuous
	and discontinuous transitions controlled by a single microscopic rule.
	(The precise sense in which the construction is ``Hamiltonian-free'' is
	made explicit in Sec.~\ref{sec:discussion}.)
	The model setup is illustrated in Fig.~\ref{fig:schematic}.
	The framework is organized in three logical layers.

	\begin{table}[b]
		\caption{Provenance of the main results.}
		\label{tab:provenance}
		\footnotesize
		\setlength{\tabcolsep}{2pt}
		\renewcommand{\arraystretch}{1.25}
		\begin{tabular}{@{}L{0.50\columnwidth}L{0.16\columnwidth}L{0.26\columnwidth}@{}}
			\toprule
			Result & Ref.~\cite{Fenoaltea2023} & This work \\
			\midrule
			Critical noise $\eta_c=\tfrac12-\tfrac1{2m}$ & yes & --- \\
			$m=2$ closed form & yes & new derivation \\
			Discontinuous transition ($f<1/2$) & yes & field reading; ordinary (not tri-) critical point \\
			arctanh fixed-point equation & --- & yes \\
			Log-odds / inference reading & --- & yes \\
			Explicit Landau-like coefficients & --- & yes \\
			Log-time relaxation scaling + MC & --- & yes \\
			Martingale convergence; quasipotential & --- & yes (standard methods) \\
			\bottomrule
		\end{tabular}
	\end{table}
	
	\textbf{Layer 1: Core theory (Sections~\ref{sec:model}--\ref{sec:beyond}).}
	The key insight is that the order-parameter dynamics can be expressed as a
	gradient flow on logarithmic time $\tau = \ln N$, with the fixed-point
	structure encoded in the compact equation
	\begin{equation}
		\arctanh(\phi^*) = m \cdot \arctanh(\alpha\phi^*)
		\label{eq:eigen_intro}
	\end{equation}
	(for the symmetric case $f=1/2$), where $\phi\in[-1,1]$ is the group
	polarization and $\alpha \equiv 1-2\eta\in[0,1]$ is the evaluation reliability.
	From this equation we construct a Landau-like effective potential with
	coefficients explicit in $m$ and $\alpha$, provide a detailed structural
	comparison with the mean-field Ising model, and analyze features that
	have no equilibrium counterpart: the $m=1$ anomaly, dictatorship and
	preferential-attachment benchmarks, history-dependent discontinuous
	transitions, and the cusp (spinodal) structure of the critical point at
	$f=1/2$.
	
	\textbf{Layer 2: Mathematical foundations (Sections~\ref{sec:clt}--\ref{sec:rg}).}
	We trace the microscopic origin of criticality to the correlations among
	verdicts induced by the shared candidate, prove almost-sure convergence to
	the stable fixed points via a P\'{o}lya-urn martingale construction, and
	show that the deterministic growth flow admits a renormalization-group (RG)
	reading with the group size $N$ as the scale parameter.
	
	\textbf{Layer 3: Complementary perspectives (Sections~\ref{sec:stoch_thermo}--
		\ref{sec:dp}).}
	We extend the analysis to a trajectory measure of irreversibility,
	information geometry, and a conjectured connection to absorbing-state
	physics in spatially embedded, zero-noise generalizations.
	
	The central thread running through the entire paper is the nested arctanh
	structure, illustrated in Fig.~\ref{fig:ising_comp}, which distinguishes
	this growth process fundamentally from the Ising model despite their shared
	universality class.
	
	We conclude with a summary of the complete paradigm and open directions
	for future work (Section~\ref{sec:discussion}).
	Technical details are in Appendices~\ref{app:noise}--\ref{app:rg_scaling}.
	
	
	\section{Model reformulation: Order parameter and deterministic flow}
	\label{sec:model}
	
	\subsection{Microscopic rules}
	
	We consider a group growing from $N_0$ founding members, all of type $+1$
	(fit for the group).
	At each discrete step, a candidate is drawn with prior probability $f$ of
	being type $+1$ and $1-f$ of being type $-1$ (unfit).
	The candidate is evaluated by $m$ group members, chosen uniformly at random
	\emph{with replacement} from the $N$ current members (the uniform case of
	Ref.~\cite{Fenoaltea2023}); the evaluators are therefore independent
	samples of the composition.
	[For sampling without replacement the unanimity probability acquires
	hypergeometric corrections of $\mathcal{O}(1/N)$, which vanish in the
	limit considered throughout; with replacement, all expressions below are
	exact at every $N$.]
	Each evaluator $i$ independently returns a positive or negative verdict:
	\begin{equation}\label{eq:micro}
		\mathbb{P}(\text{positive} \mid \sigma_i, c_j) =
		\begin{cases}
			1-\eta, & \sigma_i = c_j, \\[4pt]
			\eta,   & \sigma_i \neq c_j,
		\end{cases}
	\end{equation}
	where $\eta \in [0, \tfrac12]$ is the evaluation noise.
	The candidate is admitted if and only if all $m$ evaluators return a
	positive verdict (unanimous consensus).
	
	Let $n_+(N)$ and $n_-(N)$ be the numbers of $+1$ and $-1$ members when the
	group size is $N$, with $n_+ + n_- = N$ and $n_+(N_0) = N_0$, $n_-(N_0) = 0$.
	
	\subsection{Order parameter}
	
	We introduce the polarization
	\begin{equation}\label{eq:phi_def}
		\begin{split}
			\phi(N)
			&\equiv \frac{n_+(N) - n_-(N)}{n_+(N) + n_-(N)}  \\
			&= 2C(N)-1 \in [-1,1],
		\end{split}
	\end{equation}
	where $C(N)=n_+(N)/N$ is the cohesion.
	Initially $\phi(N_0)=1$ (all fit members).
	
	\subsection{Single-evaluator probability}
	
	A randomly chosen group member is $+1$ with probability $C=(1+\phi)/2$.
	Defining the evaluation reliability
	\begin{equation}\label{eq:alpha_def}
		\alpha \equiv 1 - 2\eta \;\in\; [0,1],
	\end{equation}
	the probability that a random evaluator approves a $+1$ candidate is
	\begin{equation}\label{eq:p_plus}
		p_+(\phi) = C(1-\eta) + (1-C)\eta = \tfrac{1}{2}\big[1 + \alpha\phi\big].
	\end{equation}
	Similarly, $p_-(\phi) = \tfrac{1}{2}[1 - \alpha\phi]$.
	The parameter $\alpha$ plays a role loosely analogous to the inverse
	temperature $\beta = 1/(k_B T)$ in the Ising model, in the limited sense
	of a control parameter for order; the analogy is structural rather than
	thermodynamic, as $\alpha$ enters the verdict probabilities linearly
	rather than through a Boltzmann factor.
	
	\subsection{Deterministic flow on logarithmic time}
	
	\subsubsection{Unconditional admission probabilities}

	Each candidate admission is the composition of two independent random events.
	First, the candidate's type is drawn from the prior: $+1$ (fit) with
	probability $f$, $-1$ (unfit) with probability $1-f$.
	Second, given the candidate's type $c_j\in\{+1,-1\}$, each of the $m$
	evaluators independently judges the candidate.  Equation~\eqref{eq:micro}
	gives the probability that an individual evaluator approves: $1-\eta$ when
	the evaluator's type matches the candidate's, $\eta$ when it does not.
	Averaging over the random choice of evaluators from the group yields
	the single-evaluator approval probabilities $p_\pm(\phi)$ of
	Eq.~\eqref{eq:p_plus}.  Since the $m$ evaluators are chosen independently
	and each must approve, the probability that all $m$ unanimously approve
	a candidate of type $\pm 1$ is $[p_\pm(\phi)]^m$.

	Multiplying the type-drawing probability by the unanimous-approval
	probability gives the \emph{unconditional} probability that an arriving
	candidate is of type $\pm 1$ \emph{and} is admitted:
	\begin{align}
	    \mathcal{P}_+(\phi) &= f \cdot [p_+(\phi)]^m
	                        = f\left(\frac{1+\alpha\phi}{2}\right)^{\!m},
	    \label{eq:P_plus}\\[4pt]
	    \mathcal{P}_-(\phi) &= (1-f) \cdot [p_-(\phi)]^m
	                        = (1-f)\left(\frac{1-\alpha\phi}{2}\right)^{\!m}.
	    \label{eq:P_minus}
	\end{align}
	The total probability that \emph{any} arriving candidate is admitted---the
	normalization factor for the admission process---is the sum of these two
	terms:
	\begin{equation}\label{eq:Z_admission}
	    Z(\phi) \equiv \mathcal{P}_+(\phi) + \mathcal{P}_-(\phi).
	\end{equation}
	Physically, $1/Z(\phi)$ is the expected number of candidate arrivals between
	two successive admissions when the group is in state $\phi$.

	\subsubsection{Conditional admission probability \texorpdfstring{$Q_+(\phi)$}{Q_+(\phi)}}

	The crucial dynamical quantity is the probability that an \emph{admitted}
	member---not an arriving candidate---is of type $+1$.
	By the definition of conditional probability,
	\begin{equation}\label{eq:Q_plus_derivation}
	    \begin{split}
	    Q_+(\phi)
	    &\equiv \mathbb{P}\bigl(\text{admitted member is }+1 \;\big|\;
	             \text{admission occurs}\bigr) \\
	    &= \frac{\mathcal{P}_+(\phi)}
	           {\mathcal{P}_+(\phi) + \mathcal{P}_-(\phi)}
	     = \frac{\mathcal{P}_+(\phi)}{Z(\phi)}.
	    \end{split}
	\end{equation}
	Substituting Eqs.~\eqref{eq:P_plus}--\eqref{eq:P_minus} into
	Eq.~\eqref{eq:Q_plus_derivation} and canceling the common factor
	$(1/2)^m$ from numerator and denominator yields the compact form
	\begin{equation}\label{eq:Q_plus}
	    Q_+(\phi)
	    = \frac{f(1+\alpha\phi)^m}
	           {f(1+\alpha\phi)^m + (1-f)(1-\alpha\phi)^m}.
	\end{equation}
	For the symmetric case $f=1/2$, this reduces to
	\begin{equation}\label{eq:Q_plus_sym}
	    Q_+(\phi)
	    = \frac{(1+\alpha\phi)^m}
	           {(1+\alpha\phi)^m + (1-\alpha\phi)^m}.
	\end{equation}

	\subsubsection{Choice of clock: why group size $N$ is the natural time variable}

	Equation~\eqref{eq:Q_plus} is a conditional probability: it conditions on
	the event of admission.
	In this paper, all dynamics is parametrized by the group size $N$, so each
	elementary step corresponds to one \emph{admitted} member.
	Rejected candidates produce no change in $\phi$ and no increment in $N$, and
	are therefore invisible in this parametrization.
	The fixed points, their stability, the bias field $h$, and the phase
	diagram are invariant under the change of clock; the relaxation-rate
	exponents and the form of the noise are clock dependent.
	All jump moments below are consequently conditioned on admission, and the
	factor $Z(\phi)$ never appears in the drift or the noise intensity---a
	significant technical simplification.

	If one instead parametrized the process by the number of candidate arrivals,
	rejected candidates would appear as null moves $\Delta\phi=0$, and the
	moments would acquire factors of $Z(\phi)$ reflecting the random number of
	rejections between successive admissions.
	Both parametrizations describe the same physical process and yield the same
	fixed-point structure; we use the growth clock throughout because it leads to
	cleaner expressions.
	\ref{app:noise} provides the explicit moment formulas for both
	clocks.
	
	\subsubsection{Drift in logarithmic time}

	Each admission step changes the order parameter by a discrete amount.
	To compute it, let $\delta\in\{0,1\}$ indicate the type of the admitted
	member: $\delta=1$ for $+1$, $\delta=0$ for $-1$.
	The conditional probability of each outcome is
	$\E[\delta\mid\phi]=Q_+(\phi)$ by definition.

	When a $+1$ member is admitted, $n_+\to n_++1$, $N\to N+1$, and
	\begin{equation*}
	    \phi' = \frac{(n_++1)-n_-}{N+1}
	          = \frac{2(n_++1)-(n_++1+n_-)}{N+1}
	           = \frac{2n_+-N + 1}{N+1}.
	\end{equation*}
	When a $-1$ member is admitted, $n_-\to n_-+1$, $N\to N+1$, and
	\begin{equation*}
	    \phi' = \frac{n_+-(n_-+1)}{N+1}
	          = \frac{2n_+-(n_++n_-+1)}{N+1}
	          = \frac{2n_+-N - 1}{N+1}.
	\end{equation*}
	Using $\phi N = 2n_+-N$, both cases are compactly written as
	\begin{equation}\label{eq:discrete_update}
	    \phi'
	    = \frac{\phi N + 2\delta - 1}{N+1},
	\end{equation}
	where $2\delta-1$ maps the Bernoulli variable $\delta\in\{0,1\}$ to the
	effect of the admitted type: $+1$ when $\delta=1$, $-1$ when $\delta=0$.

	The exact one-step increment is therefore
	\begin{equation}\label{eq:increment_exact}
	    \Delta\phi \equiv \phi' - \phi
	    = \frac{\phi N + 2\delta - 1 - \phi(N+1)}{N+1}
	    = \frac{2\delta - 1 - \phi}{N+1}.
	\end{equation}
	Its conditional expectation---the deterministic component of the
	dynamics---is
	\begin{align}
	    \E[\Delta\phi\mid\phi]
	    &= \frac{2\E[\delta\mid\phi]-1-\phi}{N+1}
	     = \frac{2Q_+(\phi)-1-\phi}{N+1}
	    \label{eq:increment_mean}.
	\end{align}

	Equation~\eqref{eq:increment_mean} has a transparent physical interpretation.
	The numerator has three contributions:
	\begin{itemize}[nosep]
	    \item $2Q_+(\phi)$ is the expected pull toward $+1$: each admitted
	          $+1$ member contributes $+1$ to $n_+-n_-$, and the factor $2$
	          arises because $n_+-n_- = 2n_+-N$ grows by $2\delta-1$ per step.
	    \item $-1$ is the constant baseline offset: even if $\phi=0$, a single
	          admission shifts the polarization from zero by $\pm 1/N$.
	    \item $-\phi$ is the dilution term: as $N$ grows, the existing
	          polarization is spread over a larger denominator, so maintaining
	          $\phi$ requires a net influx of the majority type.
	\end{itemize}
	The net effect is $\mathcal{O}(1/N)$---each admission changes the group
	composition by an ever smaller fraction as the group grows.

	The decisive step is a change of clock.
	Define logarithmic time $\tau \equiv \ln(N/N_0)$.
	A unit increment $N\to N+1$ advances $\tau$ by
	\begin{equation*}
	    \Delta\tau = \ln(N+1) - \ln N
	                = \ln\!\left(1+\frac{1}{N}\right)
	                = \frac{1}{N} + \mathcal{O}(N^{-2}).
	\end{equation*}
	Dividing the expected increment by $\Delta\tau$,
	\begin{equation}\label{eq:drift_derivation_detail}
	    \begin{split}
	    \frac{\E[\Delta\phi\mid\phi]}{\Delta\tau}
	    &= \frac{\bigl(2Q_+(\phi)-1-\phi\bigr)/(N+1)}
	            {1/N + \mathcal{O}(N^{-2})} \\
	    &= \bigl(2Q_+(\phi)-1-\phi\bigr)\cdot\frac{N}{N+1}
	      + \mathcal{O}(N^{-1}),
	    \end{split}
	\end{equation}
	which converges to a finite, nonzero limit as $N\to\infty$:
	\begin{equation}\label{eq:drift}
	    \frac{\diff\phi}{\diff\tau}
	    \equiv \mu(\phi)
	    = 2Q_+(\phi) - (1+\phi).
	\end{equation}

	The cancellation of the two $1/N$ factors is what makes the logarithmic
	clock natural.
	In the linear clock $\diff/\diff N$, the drift would be $\mu(\phi)/N$ and
	the equation non-autonomous, although its zeros---and hence the limiting
	compositions---are unchanged.
	On the logarithmic clock, $\diff\tau = \diff N/N$ exactly compensates this
	attenuation, revealing the autonomous, scale-invariant dynamics
	$\diff\phi/\diff\tau = \mu(\phi)$.
	Physically, adding one member to a group of $N=10$ alters the composition by
	$\sim 10\%$; adding one to $N=10^6$ alters it by $\sim 10^{-6}$.
	The logarithmic clock absorbs this scale dependence: equal increments of
	$\tau$ correspond to comparable fractional changes regardless of absolute
	size.
	The logarithmic clock does not create the transition---the fixed points are
	determined by $\mu(\phi^*)=0$ in any parametrization---but it renders the
	flow autonomous and the fixed-point and scaling analysis transparent; the
	formulation of this paper rests on this change of variable.
	
	For $f=1/2$:
	\begin{equation}\label{eq:mu_sym}
		\mu(\phi)
		=
		2\frac{(1+\alpha\phi)^m}
		{(1+\alpha\phi)^m + (1-\alpha\phi)^m}
		-(1+\phi).
	\end{equation}
	
	\subsubsection{Effective potential as a gradient flow}

	Equation~\eqref{eq:drift} is a first-order autonomous ODE for a single
	real variable $\phi\in[-1,1]$.
	For \emph{any} one-dimensional ODE $\dot x = f(x)$, the fundamental theorem
	of calculus guarantees that one can define a potential function
	$V(x) = -\int^{x} f(y)\,\diff y$ such that $\dot x = -\diff V/\diff x$.
	This is purely a mathematical construction---no physical postulate is
	required.

	Applying this to the drift $\mu(\phi)$ and choosing the initial condition
	$\phi=1$ (all founding members are $+1$) as the reference point, we define
	\begin{equation}\label{eq:F_eff_definition}
	    \mathcal{F}_{\eff}(\phi)
	    \equiv -\int_{1}^{\phi} \mu(y)\,\diff y,
	\end{equation}
	which satisfies $\diff\phi/\diff\tau = -\diff\mathcal{F}_{\eff}/\diff\phi$
	by construction.
	We call $\mathcal{F}_{\eff}$ the (Landau-like) \emph{effective potential};
	the suggestive notation is retained only by analogy with equilibrium free
	energies, to which it is compared---and contrasted---below.
	Along any deterministic trajectory,
	\begin{equation*}
	    \frac{\diff\mathcal{F}_{\eff}}{\diff\tau}
	    = \frac{\diff\mathcal{F}_{\eff}}{\diff\phi}\cdot
	      \frac{\diff\phi}{\diff\tau}
	    = -\mu(\phi)^2 \le 0,
	\end{equation*}
	so $\mathcal{F}_{\eff}$ decreases monotonically and can only stop at points
	where $\mu(\phi^*)=0$.
	It is therefore a Lyapunov function for the deterministic dynamics, and its
	local minima coincide exactly with the stable fixed points of the growth
	process.

	We emphasize that $\mathcal{F}_{\eff}$ is \emph{not} a free energy in the
	equilibrium sense.
	In equilibrium statistical mechanics, the free energy $F=-k_BT\ln Z$ is
	\emph{fundamental}: it is derived from the Hamiltonian via the partition
	function, and all thermodynamic observables follow from its derivatives.
	Here, $\mathcal{F}_{\eff}$ is \emph{derived}: it is the integral of the
	drift field $\mu(\phi)$, which itself follows from the probabilistic
	admission rule $Q_+(\phi)$ of Eq.~\eqref{eq:Q_plus}.
	One cannot take a temperature derivative of $\mathcal{F}_{\eff}$ because
	there is no temperature, only the evaluation noise $\eta$, which enters
	$\mu(\phi)$ through $\alpha=1-2\eta$ in a fundamentally different way than
	temperature enters the Boltzmann factor.
	The existence of $\mathcal{F}_{\eff}$ reflects only the one-dimensional
	nature of the order parameter, not an underlying equilibrium principle.
	Nevertheless, its minima correctly predict the stable phases, and its
	expansion near criticality yields the Landau coefficients derived below.
	Three distinct objects should not be conflated in what follows:
	(i)~$\mathcal{F}_{\eff}$, the Lyapunov potential of the deterministic
	drift, which exists for any one-dimensional flow;
	(ii)~the Freidlin--Wentzell quasipotential $S(\phi)$ of
	Sec.~\ref{sec:quasipotential}, which controls the $\mathcal{O}(1/N)$
	fluctuations and, because the noise $\sigma^2(\phi)$ is state dependent,
	is \emph{not} proportional to $\mathcal{F}_{\eff}$;
	and (iii)~an equilibrium free energy derived from a partition function,
	which does not exist here.
	
	\section{The arctanh representation}
	\label{sec:arctanh}
	
	\subsection{Derivation}
	
	Setting $\mu(\phi^*)=0$ in Eq.~\eqref{eq:mu_sym} and cross-multiplying:
	\begin{equation}\label{eq:fp_raw_sym}
		\frac{(1+\alpha\phi^*)^m}
		{(1+\alpha\phi^*)^m+(1-\alpha\phi^*)^m}
		=
		\frac{1+\phi^*}{2}.
	\end{equation}
	\begin{equation}\label{eq:fp_cross}
		2(1+\alpha\phi^*)^m
		= (1+\phi^*)(1+\alpha\phi^*)^m
		+ (1+\phi^*)(1-\alpha\phi^*)^m .
	\end{equation}
	\begin{equation}\label{eq:fp_balanced}
		(1-\phi^*)(1+\alpha\phi^*)^m = (1+\phi^*)(1-\alpha\phi^*)^m .
	\end{equation}
	Rearranging Eq.~\eqref{eq:fp_balanced} gives
	\begin{equation}\label{eq:fp_ratio_sym}
		\frac{1+\phi^*}{1-\phi^*} = \left(\frac{1+\alpha\phi^*}{1-\alpha\phi^*}\right)^{\!m}.
	\end{equation}
	
	Taking $\frac12\ln(\cdot)$ of both sides yields the central result:
	\begin{equation}\label{eq:arctanh_eigen}
		\arctanh(\phi^*) = m \cdot \arctanh(\alpha\phi^*).
	\end{equation}
	
	For $f\neq 1/2$, defining $h \equiv \frac12\ln[f/(1-f)]$:
	\begin{equation}\label{eq:arctanh_eigen_asym}
		\arctanh(\phi^*) = h + m \cdot \arctanh(\alpha\phi^*).
	\end{equation}
	Here $h$ is the exact analog of an external field.  When $f>1/2$, $h>0$
	and the candidate pool favors the founding $+1$ type.  When $f<1/2$, $h<0$
	and the candidate pool opposes the initial condition.  The asymmetric
	fixed-point equation before taking the logarithm is
	\begin{equation}\label{eq:fp_ratio_asym}
		\frac{1+\phi^*}{1-\phi^*}
		=
		\frac{f}{1-f}
		\left(\frac{1+\alpha\phi^*}{1-\alpha\phi^*}\right)^{m}.
	\end{equation}
	\subsection{Physical meaning: log-odds and self-consistent inference}
	\label{sec:bayes}

	Before listing formal consequences, we identify what the arctanh variable
	\emph{is}.
	For a candidate of type $c=\pm1$, a single evaluator returns a positive
	verdict with probability $p_{\pm}(\phi)=\tfrac12(1\pm\alpha\phi)$
	[Eq.~\eqref{eq:p_plus}].
	The log-likelihood ratio that one positive verdict carries about the
	candidate's type is therefore
	\begin{equation}\label{eq:llr}
		\ln\frac{p_+(\phi)}{p_-(\phi)}
		= \ln\frac{1+\alpha\phi}{1-\alpha\phi}
		= 2\arctanh(\alpha\phi).
	\end{equation}
	arctanh is the \emph{log-odds variable} of the noisy evaluation process,
	with the composition $\phi$ entering as the population of evaluators,
	attenuated by the reliability $\alpha$.
	Because the $m$ verdicts are conditionally independent given the candidate,
	unanimity adds $m$ such terms, and the prior $f$ contributes its own
	log-odds $2h=\ln[f/(1-f)]$.
	Bayes' rule then gives, for the conditional probability
	$Q_+$ of Eq.~\eqref{eq:Q_plus},
	\begin{equation}\label{eq:bayes_posterior}
		\ln\frac{Q_+(\phi)}{1-Q_+(\phi)}
		= 2h + 2m\,\arctanh(\alpha\phi),
	\end{equation}
	i.e.\ $Q_+$ is exactly the Bayesian posterior probability that the
	admitted member is fit, given $m$ unanimous verdicts and the prior $f$.
	Meanwhile the composition itself has log-odds
	$\arctanh(\phi)=\tfrac12\ln(n_+/n_-)$.
	The fixed-point condition $Q_+(\phi^*)=(1+\phi^*)/2$ therefore states
	that the group's composition log-odds reproduces the posterior log-odds of
	its own admissions---Eq.~\eqref{eq:arctanh_eigen_asym} is the
	self-consistency condition of an inference loop in which the group
	evaluates candidates using itself as the reference population.

	In this reading the phase transition acquires an information-theoretic
	meaning.
	Near $\phi=0$ the posterior log-odds responds to the composition log-odds
	with gain $\diff[m\arctanh(\alpha\phi)]/\diff[\arctanh\phi]\to
	m\alpha$: each evaluator transmits the composition through a noisy binary
	channel of reliability $\alpha$, and consensus multiplies the number of
	independent channel uses.
	Order persists when the loop gain exceeds unity, $m\alpha>1$, which is
	precisely the critical condition $\alpha_c=1/m$ of
	Eq.~\eqref{eq:alpha_c_box}; the $m=1$ anomaly is the statement that a
	single noisy reading of the composition always has gain $\alpha<1$ and can
	never regenerate it.
	The growth model is thus an exact mean-field theory of \emph{self-consistent
	Bayesian inference}, and the arctanh representation is not an algebraic
	coincidence but the natural coordinate in which evidence is additive.
	We stress that ``Bayesian'' here describes the mathematical structure of
	the admission rule, not the agents: the evaluators are simple noisy
	voters, and the posterior of Eq.~\eqref{eq:bayes_posterior} emerges
	mechanically from independent verdicts combined by unanimity.

	\subsection{Consequences of the arctanh representation}

	Equation~\eqref{eq:arctanh_eigen} is not merely a cosmetic rewriting of the
	fixed-point condition---it reorganizes the mathematics of the model so that
	questions previously requiring numerical work become answerable by
	elementary algebra.
	Five consequences stand out.

	\textbf{1.  Linearization in arctanh space.}
	The multiplicative effect of the $m$ independent evaluators,
	$\bigl(\frac{1+\alpha\phi^*}{1-\alpha\phi^*}\bigr)^m$, becomes a simple
	multiplication: $m\cdot\arctanh(\alpha\phi^*)$.
	This is analogous to how a Fourier transform converts convolution into
	multiplication---a change of basis turns a nonlinear operation into a linear
	one, at the cost of working with $\arctanh(\phi)$ rather than $\phi$ itself.
	In the asymmetric case, the prior bias $f$ and the
	evaluator count $m$, which are entangled in the original rational
	form of the drift, decouple into two additive terms:
	\begin{equation*}
	    \arctanh(\phi^*) = h + m\cdot\arctanh(\alpha\phi^*),
	\end{equation*}
	where $h=\frac12\ln[f/(1-f)]$ is the exact analog of an external magnetic
	field.
	This means that once the symmetric ($f=1/2$) theory is solved, the full
	asymmetric phase diagram is obtained by intersecting the \emph{same} function
	$\Phi(\phi)=\arctanh(\phi)-m\cdot\arctanh(\alpha\phi)$ with a
	horizontal line at height $h$---a geometric construction with no
	counterpart in the original variables.

	\textbf{2.  Fixed-point topology becomes algebraic.}
	In the original drift $\mu(\phi^*)=0$, determining the number and stability of
	fixed points for general $m$ requires numerical root-finding.
	In the arctanh form, define $\Phi(\phi)=m\cdot\arctanh(\alpha\phi)-\arctanh(\phi)$.
	Since $\Phi$ is odd for $f=1/2$, $\phi=0$ is always a fixed point.
	Its stability follows from $\Phi'(0)=m\alpha-1$: stable when
	$m\alpha<1$, unstable when $m\alpha>1$, with the sign change at
	$\alpha_c=1/m$ signaling the bifurcation.
	The number of nonzero fixed points is determined by the sign of the cubic term
	in the Taylor expansion of $\Phi$ at the origin---a one-line algebra exercise
	rather than a case-by-case numerical search.
	In particular, one proves that $m\ge 2$ yields a supercritical pitchfork
	bifurcation, while $m=1$ yields no bifurcation at all, without numerically
	solving any transcendental equation.

	\textbf{3.  Critical exponents from the Taylor series.}
	Expanding $\arctanh(x)=x+x^3/3+x^5/5+\cdots$ in
	Eq.~\eqref{eq:arctanh_eigen} gives the series expansion of $\Phi$
	displayed in Eq.~\eqref{eq:Phi_expansion} below.
	The fixed-point condition for general $f$ is $\Phi(\phi^*)=h$, where
	$h=\frac12\ln[f/(1-f)]$ is the external-field analog.
	Each critical exponent follows from the leading-order balance of this single
	equation near $\alpha=\alpha_c=1/m$.

	\textbf{Order-parameter exponent $\beta$:}
	For the symmetric case $h=0$, write $\alpha=\alpha_c+\varepsilon$ with
	$0<\varepsilon\ll 1$.
	Then $1-m\alpha = -m\varepsilon$, and the fixed-point equation becomes
	\begin{equation*}
	    -m\varepsilon\,\phi
	    + \frac{1-m\alpha_c^3}{3}\phi^3
	    + \mathcal{O}(\varepsilon\phi^3,\phi^5) = 0.
	\end{equation*}
	Since $1-m\alpha_c^3=1-1/m^2\neq0$ for $m\ge 2$, the cubic coefficient
	is non-vanishing at criticality.
	Factoring out $\phi$ (the $\phi=0$ trivial branch) and keeping the leading
	terms gives
	\begin{equation*}
	    \phi^2 = \frac{3m\varepsilon}{1-1/m^2} + \mathcal{O}(\varepsilon^2),
	\end{equation*}
	hence $\phi^*\propto(\alpha-\alpha_c)^{1/2}$ and $\beta=1/2$.

	\textbf{Susceptibility exponent $\gamma$:}
	In the disordered phase ($\alpha<\alpha_c$), the only fixed point at $h=0$
	is $\phi^*=0$.
	For small $h$, solving $\Phi(\phi)=h$ perturbatively around $\phi=0$ gives
	$h = \Phi'(0)\,\phi + \mathcal{O}(\phi^3) = (1-m\alpha)\phi +
	\mathcal{O}(\phi^3)$.
	The linear susceptibility is therefore
	\begin{equation*}
	    \chi \equiv \left.\frac{\partial\phi}{\partial h}\right|_{h=0}
	        = \frac{1}{\Phi'(0)}
	        = \frac{1}{1-m\alpha}
	        \propto |\alpha-\alpha_c|^{-1},
	\end{equation*}
	giving $\gamma=1$.

	\textbf{Critical-isotherm exponent $\delta$:}
	Exactly at $\alpha=\alpha_c$, the linear term $1-m\alpha$ vanishes.
	The equation of state reduces to $h = \frac{1-m\alpha_c^3}{3}\phi^3 +
	\mathcal{O}(\phi^5)$.
	Inverting gives $\phi \propto h^{1/3}$, hence $\delta=3$.

	Each of these exponents emerges directly from the Taylor coefficients of
	$\Phi(\phi)$---no numerical fitting, no phenomenological parametrization,
	and no approximation beyond the local expansion near the critical point.

	\textbf{4.  Landau coefficients are derived, not fitted.}
	Using the arctanh form, the drift is compactly written as
	$\mu(\phi)=\tanh(m\cdot\arctanh(\alpha\phi))-\phi$.
	Expanding in powers of $\phi$ and integrating yields the Landau-like
	potential with coefficients
	\begin{equation*}
	\begin{gathered}
	    r = 1-m\alpha,\qquad
	    u = \frac{m\alpha^3}{3}(m^2-1),\\
	    v = -\frac{m\alpha^5}{15}(2m^4-5m^2+3),
	\end{gathered}
	\end{equation*}
	expressed entirely in the microscopic parameters $m$ and $\alpha$.
	The quartic coefficient $u\propto(m^2-1)$ makes transparent why $m=1$ is
	degenerate ($u=0$, no phase transition) while $m\ge 2$ supports a continuous
	transition ($u>0$).
	In standard Landau theory, these coefficients are phenomenological parameters
	to be determined experimentally; here they are closed-form consequences of the
	admission rule.

	\textbf{5.  Direct structural comparison with the Ising model.}
	The mean-field Ising fixed-point equation
	$m_I=\tanh(\beta J q\,m_I+\beta H)$ (we write $m_I$ for the Ising
	magnetization, to avoid confusion with the evaluator number $m$)
	rewrites as $\arctanh(m_I)=\beta J q\,m_I+\beta H$.
	Placing this next to $\arctanh(\phi^*)=m\cdot\arctanh(\alpha\phi^*)+h$
	reveals a fundamental structural difference: the Ising relation is
	\emph{linear in the bare order parameter}, whereas the growth model is
	\emph{linear in the arctanh of the attenuated order parameter}.
	Both belong to the mean-field Ising universality class, but they reach it
	through mathematically distinct routes---the nested arctanh structure is
	unique to the growth process and has no Ising counterpart.
	This distinction, summarized in Table~\ref{tab:ising_comparison}, would be
	invisible without the arctanh representation.

	\subsection{Example: closed-form solution for \texorpdfstring{$m=2$}{m=2}}

	For $m=2$, Eq.~\eqref{eq:arctanh_eigen} admits the closed-form solution
	$\phi^* = \sqrt{2\alpha-1}/\alpha$ for $\alpha>1/2$, obtained via the
	double-angle identity $\arctanh(2x/(1+x^2)) = 2\arctanh(x)$:
	\begin{equation}\label{eq:m2_closed}
	    \phi^*=\pm\frac{\sqrt{2\alpha-1}}{\alpha}.
	\end{equation}
	In terms of the original noise parameter $\eta$, this is
	$\phi^*=\pm\sqrt{1-4\eta}/(1-2\eta)$, recovering the result of
	Ref.~\cite{Fenoaltea2023} by a considerably more direct route.

	\section{Phase-transition landscape and comparison with the Ising model}
	\label{sec:landau}
	
	\subsection{Pitchfork bifurcation and critical exponents}
	
	Define $\Phi(\phi) \equiv \arctanh(\phi) - m\cdot\arctanh(\alpha\phi)$,
	so fixed points satisfy $\Phi(\phi^*)=0$ ($f=1/2$).
	Expanding $\arctanh(x) = x + x^3/3 + x^5/5 + \mathcal{O}(x^7)$:
	\begin{align}
		\Phi(\phi) &= (1 - m\alpha)\phi
		+ \frac{1 - m\alpha^3}{3}\phi^3
		+ \frac{1 - m\alpha^5}{5}\phi^5
		+ \mathcal{O}(\phi^7).
		\label{eq:Phi_expansion}
	\end{align}
	Factoring out the trivial solution isolates the nonzero branches:
	\begin{equation}\label{eq:Phi_factor}
		0 = 1-m\alpha
		+ \frac{1-m\alpha^3}{3}\phi^2
		+ \frac{1-m\alpha^5}{5}\phi^4
		+ \mathcal{O}(\phi^6).
	\end{equation}
	
\begin{figure}[htbp]
		\centering
		\includegraphics[width=\columnwidth]{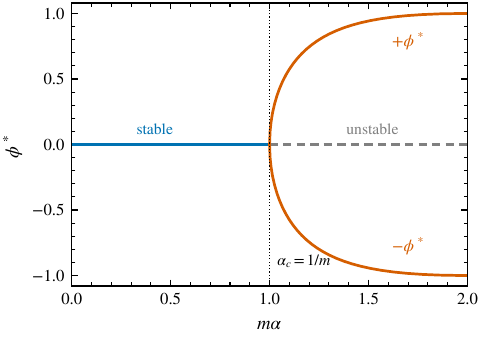}
		\caption{Supercritical pitchfork bifurcation for $f=1/2$, $m=2$.
			Solid: stable fixed points; dashed: unstable.
			$\alpha_c=1/m$ marks the critical point.}
		\label{fig:bifurcation}
	\end{figure}

	The linear stability of $\phi^*=0$ is governed by
	\begin{equation}\label{eq:mu_prime_0}
		\mu'(0) = m\alpha - 1.
	\end{equation}
	Thus the critical point is
	\begin{equation}\label{eq:alpha_c_box}
		\alpha_c = \frac{1}{m},\qquad \eta_c = \frac{1}{2} - \frac{1}{2m},
	\end{equation}
	with $\phi^* \propto (\alpha-\alpha_c)^{1/2}$, giving $\beta = 1/2$.
	More explicitly, retaining the leading nonzero term in
	Eq.~\eqref{eq:Phi_factor} gives
	\begin{equation}\label{eq:phi_scaling_detailed}
		\phi^2
		=
		\frac{3(m\alpha-1)}{1-m\alpha^3}.
	\end{equation}
	At $\alpha=\alpha_c=1/m$, the denominator is $1-1/m^2\neq0$ for
	$m\ge2$, so
	\begin{equation}\label{eq:beta_scaling_detailed}
		\phi^*
		\simeq
		\sqrt{\frac{3m}{1-1/m^2}}\,
		(\alpha-\alpha_c)^{1/2}.
	\end{equation}
	The supercritical pitchfork bifurcation is shown in Fig.~\ref{fig:bifurcation}.
	
	\subsection{Landau expansion with explicit coefficients}
	
	The Landau expansion is obtained by expanding the drift $\mu(\phi)$ as a
	power series and then integrating.  We begin from the identity
	\begin{equation}\label{eq:mu_tanh_identity}
		\mu(\phi)
		=
		\tanh\!\big(m\,\arctanh(\alpha\phi)\big)-\phi,
	\end{equation}
	which follows directly from $Q_+(\phi)=\frac12[1+\tanh(m\,\arctanh(\alpha\phi))]$
	for $f=1/2$ and the definition $\mu(\phi)=2Q_+(\phi)-1-\phi$.
	
	To expand systematically, define the auxiliary variable
	$\theta \equiv \arctanh(\alpha\phi)$.
	Expanding $\theta$ in powers of $\phi$:
	\begin{equation}\label{eq:theta_expansion}
		\theta
		= \arctanh(\alpha\phi)
		= \alpha\phi + \frac{\alpha^3}{3}\phi^3
		+ \frac{\alpha^5}{5}\phi^5
		+ \mathcal{O}(\phi^7).
	\end{equation}
	Next, expand $\tanh(m\theta)$ as a function of $\theta$:
	\begin{equation}\label{eq:tanh_expansion}
		\tanh(m\theta)
		= m\theta - \frac{m^3}{3}\theta^3
		+ \frac{2m^5}{15}\theta^5
		+ \mathcal{O}(\theta^7).
	\end{equation}
	Substituting Eq.~\eqref{eq:theta_expansion} into Eq.~\eqref{eq:tanh_expansion}
	and collecting powers of $\phi$:
	\begin{align}
		\tanh(m\,\arctanh(\alpha\phi))
		&= m\!\left(\alpha\phi + \frac{\alpha^3}{3}\phi^3
		+ \frac{\alpha^5}{5}\phi^5\right) \notag\\
		&\quad - \frac{m^3}{3}\!\left(\alpha\phi + \frac{\alpha^3}{3}\phi^3\right)^{\!3} \notag\\
		&\quad + \frac{2m^5}{15}(\alpha\phi)^5
		+ \mathcal{O}(\phi^7) \notag\\[4pt]
		&= m\alpha\phi
		+ \frac{m\alpha^3}{3}(1-m^2)\phi^3 \notag\\
		&\quad + \frac{m\alpha^5}{15}(2m^4-5m^2+3)\phi^5
		\notag\\
		&\quad + \mathcal{O}(\phi^7).
		\label{eq:compound_expansion}
	\end{align}
	In the cubic term, the factor $m\alpha^3/3$ comes from the linear
	$\theta$-to-$\phi$ mapping, while the $-m^3\alpha^3/3$ comes from the
	cubic term in $\tanh(m\theta)$; together they produce the factor
	$(1-m^2)$, which vanishes at $m=1$---this is the Landau-level manifestation
	of the $m=1$ anomaly discussed in Sec.~\ref{sec:beyond}.
	
	Subtracting $\phi$ from Eq.~\eqref{eq:compound_expansion} gives the drift
	expansion:
	\begin{align}
		\mu(\phi)
		&= (m\alpha-1)\phi
		- \frac{m\alpha^3(m^2-1)}{3}\phi^3 \notag\\
		&\quad + \frac{m\alpha^5(2m^4-5m^2+3)}{15}\phi^5
		+ \mathcal{O}(\phi^7).
		\label{eq:drift_expansion}
	\end{align}
	
	The effective potential satisfies $\mathcal{F}_{\eff}'(\phi)=-\mu(\phi)$.
	Integrating $-\mu(\phi)$ term by term from $\phi=0$ (where $\mathcal{F}_{\eff}$
	is defined up to an additive constant) yields the Landau form:
	\begin{equation}\label{eq:landau_F}
		\mathcal{F}_{\eff}(\phi) = \mathcal{F}_0 + \frac{r}{2}\phi^2
		+ \frac{u}{4}\phi^4 + \frac{v}{6}\phi^6 + \mathcal{O}(\phi^8),
	\end{equation}
	with coefficients that are explicit functions of the microscopic parameters:
	\begin{align}
		r &= 1 - m\alpha, \label{eq:r_coeff}\\[4pt]
		u &= \frac{m\alpha^3}{3}(m^2-1), \label{eq:u_coeff}\\[4pt]
		v &= -\frac{m\alpha^5}{15}(2m^4-5m^2+3). \label{eq:v_coeff}
	\end{align}
	
	Each coefficient has a direct physical meaning.  The quadratic coefficient
	$r = 1-m\alpha$ is the inverse susceptibility: it changes sign at the
	critical point $\alpha_c=1/m$, signaling the instability of the disordered
	phase.  The quartic coefficient $u = m\alpha^3(m^2-1)/3$ is positive for all
	$m\ge 2$ and all $\alpha>0$, ensuring that the transition remains continuous
	at $f=1/2$ (the Landau functional is bounded from below by the quartic term).
	The factor $(m^2-1)$ makes precise the intuition that a single evaluator
	($m=1$) cannot generate nonlinear feedback: when $m=1$, $u=0$ identically,
	and the Landau theory degenerates to a pure Gaussian model.  The sixth-order
	coefficient $v$ stays subdominant at the continuous transition; it would
	govern the local theory only if the quartic coupling $u$ were tuned to zero
	(Sec.~\ref{sec:tricritical}), which the unanimity rule never does.
	
	These coefficients are \emph{microscopic}: no phenomenological fitting has
	been introduced.  Every term is derived from the original evaluation
	probabilities through the exact arctanh representation.
	
	For $m=1$, $u=0$ (degenerate); for $m\ge 2$, $u>0$, so the quartic term
	stabilizes the local theory at the continuous transition.  The linear
	susceptibility $\chi = 1/r \propto |\alpha-\alpha_c|^{-1}$ gives $\gamma=1$,
	and at criticality $\phi \propto h^{1/3}$ gives $\delta=3$.
	
	Beyond the linear response, the \emph{nonlinear susceptibility}
	$\chi_3 \equiv \partial^3\phi/\partial h^3|_{h=0}$ provides a sharper
	signature of the transition.
	The physical
	conjugate field is the candidate bias $h=\tfrac12\ln[f/(1-f)]$, which enters
	the fixed-point condition additively as $\Phi(\phi)=h$
	[Eq.~\eqref{eq:arctanh_eigen_asym}], so the relevant equation of state is the
	expansion of $\Phi$, Eq.~\eqref{eq:Phi_expansion},
	\begin{equation}\label{eq:eos_field}
		r\,\phi + u_\Phi\,\phi^3 + \mathcal{O}(\phi^5) = h,
		\qquad
		u_\Phi = \frac{1-m\alpha^3}{3},
	\end{equation}
	with $r=1-m\alpha$ as before.  Note that the cubic coefficient $u_\Phi$ of
	the \emph{field} equation of state differs from the Landau quartic
	coefficient $u=m\alpha^3(m^2-1)/3$ of $\mathcal{F}_{\eff}$ [the latter
	governs the response to a field conjugate to $\phi$ in the gradient-flow
	potential, not to the physical bias]; the two coincide exactly at
	criticality, $u_\Phi(\alpha_c)=u(\alpha_c)=(m^2-1)/(3m^2)$, which is why the
	critical-isotherm exponent $\delta=3$ is unambiguous.
	Differentiating Eq.~\eqref{eq:eos_field} three times and evaluating at
	$h=0$ yields
	\begin{equation}\label{eq:chi3}
		\chi_3 = -\frac{6\,u_\Phi}{r^4}
		\propto |\alpha-\alpha_c|^{-4}.
	\end{equation}
	The exponent $\gamma_3 = 4$ (independent of the amplitude) satisfies the
	mean-field scaling relation $\gamma_3 = 3\Delta - \beta = \gamma + 2\Delta$,
	where $\Delta = \beta + \gamma = 3/2$ is the gap exponent.
	Near criticality, this provides a falsifiable prediction: the polarization
	response to a small external bias $h$ is not merely amplified but becomes
	\emph{cubically nonlinear}, with
	$\phi(h) \simeq \chi h + \tfrac{1}{6}\chi_3 h^3$.
	
	The phase classification is therefore:
	\begin{itemize}[nosep]
		\item For $m=1$, the quartic coupling vanishes and the transition is pushed
		to $\alpha_c=1$ or $\eta_c=0$.  The deterministic flow is purely
		linear, so there is no finite critical noise.
		\item For $m\ge2$, $u>0$ for every $\alpha>0$, ensuring a bounded Landau
		functional and a continuous supercritical pitchfork at
		$\alpha_c=1/m$.
	\end{itemize}
	
\begin{figure}[htbp]
		\centering
		\includegraphics[width=\columnwidth]{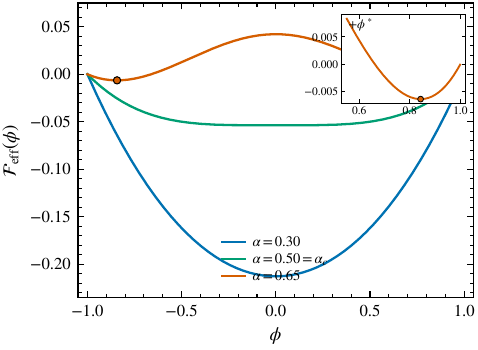}
		\caption{Landau-like effective potential $\mathcal{F}_{\eff}(\phi)$ for
			$m=2$, $f=1/2$.
			$\alpha<\alpha_c$: single well (disordered).
			$\alpha=\alpha_c$: quartic minimum (critical).
			$\alpha>\alpha_c$: double-well (ordered).}
		\label{fig:F_eff}
	\end{figure}

	For $m=2$, $\mathcal{F}_{\eff}$ integrates in closed form
	(see Fig.~\ref{fig:F_eff}):
	\begin{align}
		\mathcal{F}_{\eff}(\phi; m=2)
		&=
		-\int_1^\phi
		\frac{y(2\alpha-1-\alpha^2y^2)}
		{1+\alpha^2y^2}\,\diff y \notag\\
		&=
		\frac{\phi^2}{2}
		-\frac{1}{\alpha}\ln(1+\alpha^2\phi^2)
		+\const .
		\label{eq:F_eff_m2}
	\end{align}
	Direct differentiation verifies
	$-\diff\mathcal{F}_{\eff}/\diff\phi=\mu(\phi)$.
	
	\subsection{Structural comparison with the mean-field Ising model}
	\label{sec:ising_comparison}
	
\begin{figure*}[t]
		\centering
		\includegraphics[width=\textwidth]{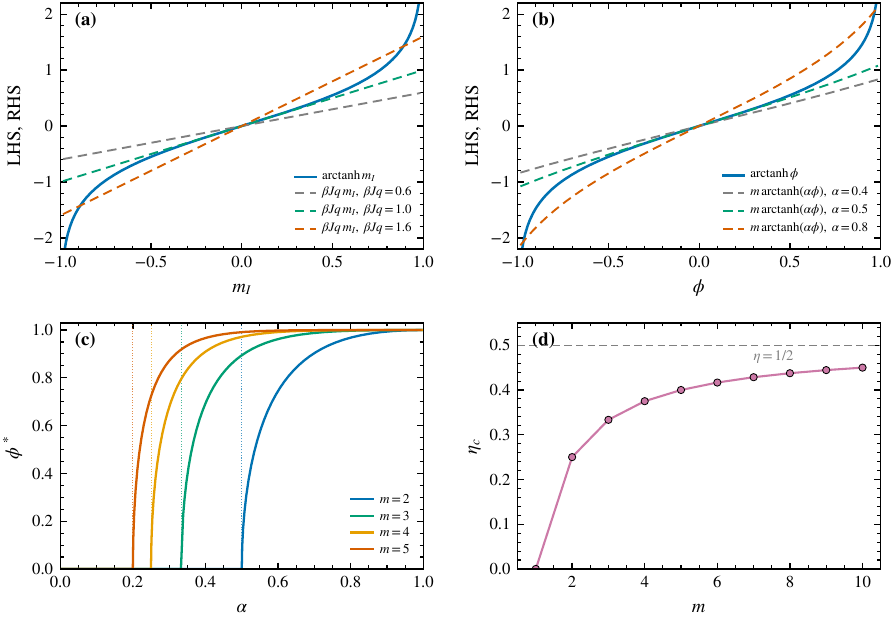}
		\caption{Structural comparison between the mean-field Ising model and the
			growing-group model.
			(a)~Ising: $\arctanh(m_I) \propto m_I$ (linear in the bare order
			parameter; $m_I$ denotes the Ising magnetization).
			(b)~Group growth: $\arctanh(\phi) \propto \arctanh(\alpha\phi)$
			(linear in the arctanh of the attenuated order parameter).
			(c)~Order parameter $\phi^*$ versus reliability $\alpha$ for various $m$.
			(d)~Critical noise $\eta_c = 1/2-1/(2m)$ versus evaluator count $m$.}
		\label{fig:ising_comp}
	\end{figure*}

	A central question is how this growth-driven phase transition relates to the
	canonical equilibrium paradigm.
	The Ising self-consistency equation
	$m_I = \tanh(\beta J q\, m_I + \beta H)$ gives
	$\arctanh(m_I) = \beta J q \cdot m_I + \beta H$, which is \emph{linear
	in $m_I$}.
	Equation~\eqref{eq:arctanh_eigen} has $\arctanh(\phi^*) \propto
	\arctanh(\alpha\phi^*)$, which is \emph{linear in $\arctanh(\alpha\phi^*)$}.
	This nested arctanh structure---illustrated in Fig.~\ref{fig:ising_comp}---
	is a consequence of the multiplicative admission process and has no Ising
	counterpart.
	The contrast can be stated physically: mean-field Ising ordering arises
	from the \emph{superposition of energies} (the molecular field adds
	contributions linearly in $m_I$), whereas growth ordering arises from the
	\emph{multiplication of likelihoods} (consensus adds log-odds,
	Sec.~\ref{sec:bayes}); the two mechanisms share exponents but not an
	equation of state.
	
	Table~\ref{tab:ising_comparison} provides the detailed item-by-item comparison.
	
	\begin{table}[htbp]
		\caption{Structural comparison between the mean-field Ising model and the
			growing-group model ($f=1/2$, $H=0$).}
		\label{tab:ising_comparison}
		\footnotesize
		\setlength{\tabcolsep}{2pt}
		\renewcommand{\arraystretch}{1.25}
		\begin{tabular}{@{}L{0.29\columnwidth}L{0.30\columnwidth}L{0.34\columnwidth}@{}}
			\toprule
			Quantity & MF Ising & Group growth \\
			\midrule
			Order parameter & $m_I\in[-1,1]$ & $\phi\in[-1,1]$ \\
			Control param.\ & $\beta J q$ & $m\alpha = m(1-2\eta)$ \\
			Fixed-point eq.\ & $m_I=\tanh(\beta J q\, m_I)$ & Eq.~(\ref{eq:fp_ratio_sym}) \\
			Arctanh form & $\arctanh(m_I)=\beta J q\,m_I$ & $\arctanh(\phi)=m\,\arctanh(\alpha\phi)$ \\
			Critical value & $(\beta J q)_c=1$ & $\alpha_c=1/m$ \\
			$\beta$ (order param.) & $1/2$ & $1/2$ \\
			$\gamma$ (suscept.) & $1$ & $1$ \\
			$\delta$ (at crit.) & $3$ & $3$ \\
			Universality & Mean-field Ising & Mean-field Ising \\
			Free energy & $F=-k_BT\ln Z$ & $\mathcal{F}_{\eff}=-\int\mu\,\diff\phi$ \\
			Fluctuation & Thermal bath & Evaluation noise \\
			Time & Real time (Model A) & $\tau=\ln N$ (RG scale) \\
			\bottomrule
		\end{tabular}
	\end{table}
	
	\begin{table}[htbp]
		\caption{Mean-field equation-of-state and log-time exponents of the
			growing-group model ($m\ge 2$, $f=1/2$).  Here $\nu_\tau$ and
			$z_\tau$ are relaxation exponents on logarithmic time
			$\tau=\ln(N/N_0)$; the all-to-all model has no spatial
			correlation length.}
		\label{tab:exponents}
		\footnotesize
		\setlength{\tabcolsep}{2pt}
		\renewcommand{\arraystretch}{1.25}
		\begin{tabular}{@{}L{0.22\columnwidth}L{0.58\columnwidth}L{0.12\columnwidth}@{}}
			\toprule
			Exponent & Definition & Value \\
			\midrule
			$\beta$  & $\phi^* \sim (\alpha-\alpha_c)^\beta$ & $1/2$ \\
			$\gamma$ & $\chi \sim |\alpha-\alpha_c|^{-\gamma}$ & $1$ \\
			$\delta$ & $\phi \sim h^{1/\delta}$ at $\alpha_c$ & $3$ \\
			$\nu_\tau$ & $\lambda_\tau^{-1} \sim |\alpha-\alpha_c|^{-\nu_\tau}$ & $1$ \\
			$z_\tau$ & $\tau_{\text{relax}} \sim \xi^{z_\tau}$ & $1$ \\
			\bottomrule
		\end{tabular}
	\end{table}
	
	The comparison reveals a nuanced relationship.
	Both models share the same static critical exponents and belong to the
	mean-field Ising universality class, but their mathematical architectures
	differ fundamentally.
	In Ising, the interaction is encoded in a Hamiltonian; in the growth model,
	it is generated by the consensus-based multiplication of independent
	evaluation probabilities.
	In Ising, the temperature controls fluctuations; in the growth model, the
	evaluation noise $\eta$ plays a dual role, simultaneously attenuating the
	signal from the group composition and providing the stochastic fluctuations
	that drive finite-size effects.
	Most importantly, the growth model exhibits several phenomena with no
	equilibrium Ising counterpart, which we analyze in the next section.
	
	\section{Beyond the Ising paradigm}
	\label{sec:beyond}
	
	\subsection{The \texorpdfstring{$m=1$}{m=1} anomaly}
	
	For $m=1$, $f=1/2$: $\mu(\phi) = (\alpha-1)\phi=-2\eta\phi$.
	With $\phi(N_0)=1$, the continuum (large-$N$) solution is
	\begin{equation}\label{eq:m1_solution}
		\phi(N) = \left(\frac{N_0}{N}\right)^{\!2\eta},\quad
		C(N) = \tfrac12 + \tfrac12\left(\frac{N_0}{N}\right)^{\!2\eta}.
	\end{equation}
	[The discrete process obeys the exact product
	$\E[\phi(N)]=\prod_{k=N_0}^{N-1}\bigl(1-\tfrac{2\eta}{k+1}\bigr)$,
	with the same asymptotics.]
	Cohesion decays as a power law; $\lim_{N\to\infty}C(N)=1/2$ for any $\eta>0$.
	This is a mean-field model with no finite critical noise---a consequence
	of the vanishing excess admission probability, as we show in
	Sec.~\ref{sec:clt}.
	The mechanism is not low dimensionality but the lack of collective
	averaging.
	For a single evaluator, every decision is exposed to order-one noise and the
	drift contains only a linear restoring force.  There is no nonlinear term that
	can stabilize a nonzero ordered branch.
	
	\subsection{Dictatorship and preferential-attachment benchmarks}
	\label{sec:benchmarks}
	
	The gradient-flow framework applies directly to the other two evaluator-selection
	mechanisms studied in Ref.~\cite{Fenoaltea2023}: dictatorship (DS) and
	preferential attachment (PA).
	Their behavior reveals why consensus-based admission ($m \ge 2$) occupies a
	unique position in the phase diagram.
	
	\subsubsection{Dictatorship (DS)}
	
	A single founder member evaluates every candidate.
	Since the evaluator is fixed to type $+1$, the single-evaluator
	approval probability does not depend on the group composition:
	\begin{equation}\label{eq:p_DS}
		p_+^{\text{DS}} = 1-\eta = \frac{1+\alpha}{2},\qquad
		p_-^{\text{DS}} = \eta = \frac{1-\alpha}{2}.
	\end{equation}
	For the symmetric case $f=1/2$, $Q_+^{\text{DS}} = 1-\eta$.
	Substituting into the drift:
	\begin{equation}\label{eq:mu_DS}
		\mu_{\text{DS}}(\phi) = 2(1-\eta) - (1+\phi) = \alpha - \phi.
	\end{equation}
	The fixed point is $\phi^*_{\text{DS}} = \alpha = 1-2\eta$, and the
	asymptotic cohesion is $C^*_{\text{DS}} = 1-\eta$.
	Unlike the UC case, $C^*_{\text{DS}} > 1/2$ for any $\eta < 1/2$---the
	dictator's fixed type acts as a permanent bias that prevents cohesion from
	decaying to the random-group level.
	There is, however, no phase transition: the drift $\mu_{\text{DS}}$ is
	linear in $\phi$, so the system always has a unique, globally stable fixed
	point without spontaneous symmetry breaking.
	
	The dictatorship limit provides a useful benchmark: whenever consensus-based
	admission surpasses $C^*_{\text{DS}}$, collective evaluation outperforms a
	perfectly informed individual.
	For $f=1/2$ and $m=2$, the condition
	$\phi^*=\sqrt{2\alpha-1}/\alpha>\alpha$ reduces to
	$\alpha^3+\alpha^2+\alpha>1$, whose unique root in $(0,1)$ is
	$\alpha\approx 0.5437$; collective evaluation therefore outperforms the
	dictator for $\eta \lesssim 0.228$.
	Closer to the critical noise $\eta_c=1/4$, the attenuated consensus
	signal falls below the dictator benchmark, even though only the
	consensus rule supports a genuine phase transition.
	
	\subsubsection{Preferential attachment (PA)}
	
	In the PA scheme, each group member $i$ carries an activity counter $k_i$,
	incremented upon each participation in an admission decision.
	The probability of selecting member $i$ as an evaluator is proportional
	to $k_i$, which biases selection toward early (statistically more fit)
	members.
	
	For $m=1$, the asymptotic cohesion decays as $C_{\text{PA}}(N) - 1/2 \sim
	N^{-\eta}$---exactly half the UC exponent $2\eta$---because PA effectively
	doubles the weight of early members.
	
	For $m \ge 2$, however, PA and UC share the \emph{identical} fixed-point
	equation in the $N \to \infty$ limit.
	Both converge to the same attractor described by
	Eq.~\eqref{eq:arctanh_eigen} \cite{Fenoaltea2023}.
	The difference lies solely in the convergence speed.
	The arctanh fixed-point equation, the critical noise $\eta_c$, the
	bifurcation structure, and all critical exponents are thus universal
	across the UC and PA selection rules.
	The dictatorship is the only mechanism that lies outside this universality
	class, precisely because it breaks the feedback loop between group
	composition and evaluation outcome.
	
	Table~\ref{tab:benchmarks} summarizes the comparison.
	
	\begin{table}[htbp]
		\caption{Comparison of the three evaluator-selection mechanisms.
			For $m \ge 2$, UC and PA share identical fixed points in the
			$N\to\infty$ limit; DS has no phase transition.}
		\label{tab:benchmarks}
		\footnotesize
		\setlength{\tabcolsep}{2pt}
		\renewcommand{\arraystretch}{1.25}
		\begin{tabular}{@{}L{0.14\columnwidth}L{0.24\columnwidth}L{0.24\columnwidth}L{0.30\columnwidth}@{}}
			\toprule
			Rule
			& $m=1$ decay
			& $m=1$ limit $C_\infty$
			& $m\ge 2$ fixed point \\
			\midrule
			UC      & $N^{-2\eta}$    & $1/2$ & arctanh equation \\
			PA & $N^{-\eta}$   & $1/2$ & same as UC \\
			DS & $N^{-1}$ (finite-size) & $1-\eta$ & N/A (no transition) \\
			\bottomrule
		\end{tabular}
	\end{table}
	
	\subsection{Discontinuous transition for \texorpdfstring{$f<1/2$}{f<1/2}}
	
\begin{figure}[htbp]
		\centering
		\includegraphics[width=\columnwidth]{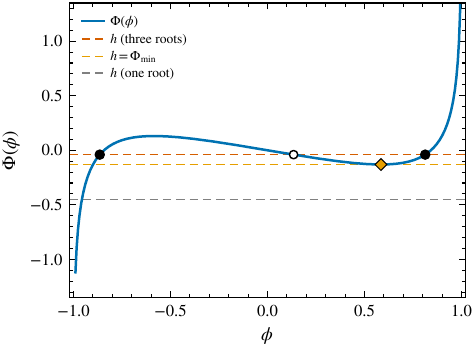}
		\caption{Saddle-node bifurcation for $f<1/2$ ($m=2$, $\alpha=0.65$).
			Small $|h|$: three roots. Critical $h=\Phi_{\min}$: saddle-node.
			Large $|h|$: only negative root.
			Filled circles: stable roots; open circle: unstable root;
			diamond: saddle-node tangency point.}
		\label{fig:saddle_node}
	\end{figure}

\begin{figure}[htbp]
		\centering
		\includegraphics[width=\columnwidth]{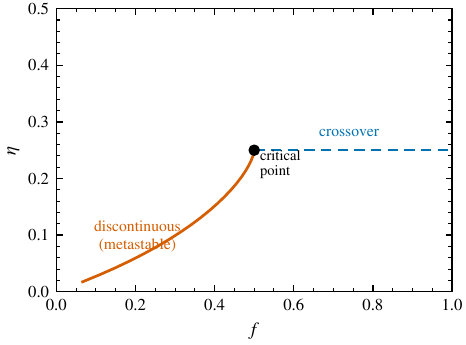}
		\caption{Phase diagram in the $(f,\eta)$ plane for $m=2$.
			Solid curve: saddle-node (spinodal) line $\eta_{\rm sn}(f)$ for
			$f<1/2$, at which the metastable cohesive branch is annihilated and
			the polarization jumps discontinuously; the jump is history
			dependent (Sec.~\ref{sec:beyond}).
			Dashed line: the locus $\eta=\eta_c$ continued to $f>1/2$, where
			the bias field $h>0$ smooths the transition into a crossover; this
			line marks no singularity and is a guide to the eye.
			Filled circle: ordinary critical point at $f=1/2$, $\eta_c=1/4$.}
		\label{fig:phase_diagram}
	\end{figure}

	When $h<0$, the saddle-node bifurcation condition is
	$\Phi(\phi_c)=h$, $\Phi'(\phi_c)=0$, giving
	\begin{equation}\label{eq:phi_c_saddle}
		\phi_c^2
		=
		\frac{m\alpha_{\rm sn}-1}{\alpha_{\rm sn}(m-\alpha_{\rm sn})},
	\end{equation}
	and an implicit equation for the saddle-node reliability
	$\alpha_{\rm sn}(f,m)$ [the metastable cohesive branch exists for
	$\alpha>\alpha_{\rm sn}(f,m)$].
	The resulting discontinuous transition
	(Figs.~\ref{fig:saddle_node} and~\ref{fig:phase_diagram})
	is history dependent rather than hysteretic in the equilibrium sense:
	members are never removed and the parameters are fixed, so no
	back-and-forth sweep protocol exists, and the jump is an
	initial-condition-dependent spinodal collapse.
	At the transition, the high-cohesion branch and the unstable saddle
	annihilate; because the initial condition is $\phi=1$, the system follows
	the metastable positive branch until this annihilation point is reached.
	The genuinely nonequilibrium feature is that the irreversible growth
	permanently records the founding condition: there is no detailed balance
	and no relaxation channel back to the disordered branch.
	
	\subsection{Ordinary critical point at \texorpdfstring{$f=1/2$}{f=1/2} and
	the cusp structure}
	\label{sec:tricritical}

	It is tempting to label the point $(f,\eta)=(1/2,\eta_c)$, where the
	wing of discontinuous transitions meets the symmetry axis, a tricritical
	point.
	Examination of the Landau coefficients shows that this is
	\emph{not} the case; the correct identification is simpler.
	The diagnostic for a tricritical point on the symmetry axis is the vanishing
	of the quartic coupling, $u(\alpha_c)=0$ \cite{Griffiths1970,Lawrie1984}.  Using the explicit coefficient
	$u=m\alpha^3(m^2-1)/3$ of Eq.~\eqref{eq:u_coeff} at the critical reliability
	$\alpha_c=1/m$,
	\begin{equation}\label{eq:u_at_crit}
		u(\alpha_c)=\frac{m\alpha_c^3(m^2-1)}{3}
		=\frac{m^2-1}{3m^2}>0
		\qquad(m\ge 2).
	\end{equation}
	The quartic coupling is \emph{strictly positive} for every $m\ge 2$, so the
	transition on the symmetry axis is always continuous and the sixth-order
	term $v$ is never required to stabilize it.  There is no tricritical point.

	The actual topology is that of an ordinary mean-field critical point with the
	standard \emph{cusp} (or spinodal) structure familiar from the Ising critical
	point.  The candidate bias $h=\tfrac12\ln[f/(1-f)]$ is a genuine
	symmetry-breaking field: $f=1/2$ is the symmetry axis $h=0$, $f>1/2$ a smooth
	crossover ($h>0$ favoring the founders), and $f<1/2$ the region $h<0$.  For
	$f<1/2$ the cohesive branch is metastable, and because the dynamics starts at
	$\phi=1$ the system follows it until the saddle-node annihilation of
	Eq.~\eqref{eq:phi_c_saddle}, producing a history-dependent discontinuous
	jump.
	The saddle-node lines $\alpha_{\rm sn}(f)$ are the two wings of a cusp whose
	tip is the critical point: as $f\to 1/2^-$ the tangency point $\phi_c\to 0$
	and $\alpha_{\rm sn}\to\alpha_c=1/m$, so the discontinuity line terminates
	exactly at the continuous transition---precisely the structure of the Ising
	critical point at the end of its $H=0$ coexistence line, not a tricritical
	point.

	The corresponding testable prediction follows from the critical isotherm.
	At $\alpha=\alpha_c$ the order parameter responds to the bias as
	$\phi\sim h^{1/3}$ ($\delta=3$), and since $h\simeq 2(f-\tfrac12)$ for $f$
	near $1/2$, the stationary polarization along the critical line scales as
	\begin{equation}\label{eq:critical_isotherm_f}
		|\phi^*(\alpha_c,f)|\;\propto\;\left|f-\tfrac12\right|^{1/3},
	\end{equation}
	with the stable root on the negative side ($\phi^*<0$) for $f<1/2$, since
	the bias favors the $-1$ type---a sharp, parameter-free signature distinct
	from the linear law that a genuine tricritical point ($\beta_t=1$) would
	predict.
	A bona fide tricritical point would require an admission rule for which
	$u(\alpha_c)=0$---for instance a majority-vote ($k$-out-of-$m$) rule, where
	the threshold $k/m$ tunes the quartic coupling through zero, realizing a
	Blume--Capel-type \cite{Blume1966,Capel1966} crossover.  We note this as a
	direction for future work rather than a property of the present
	unanimity rule.

	\section{Dynamics and finite-size effects}
	\label{sec:dynamics}
	
	\subsection{Dynamic critical exponent}
	\label{sec:dynamic_z}
	
	Near a fixed point, a small perturbation $\delta\phi$ relaxes as
	$\diff\,\delta\phi/\diff\tau = \mu'(\phi^*)\,\delta\phi$.
	At $\phi^*=0$, $\mu'(0)=m\alpha-1$.
	The logarithmic relaxation rate is $\lambda_\tau = 1 - m\alpha
	= m(\alpha_c-\alpha)$, which vanishes linearly at criticality.
	
	Transforming back to physical group size,
	$\delta\phi(N) \sim N^{-\lambda_\tau}$---power-law rather than exponential
	decay.
	The dynamic exponent, defined via $\tau_{\text{relax}} \sim \xi^z$ with
	$\xi \sim |\alpha-\alpha_c|^{-1}$, is
	\begin{equation}\label{eq:z}
		z = 1 \quad (\text{on logarithmic time } \tau).
	\end{equation}
	We stress that $z=1$ is a \emph{log-time relaxation exponent}: because
	$\tau=\ln N$ serves simultaneously as the dynamical time and as the flow
	scale, the relation $\tau_{\text{relax}}\sim\xi^{z}$ with $z=1$ holds by
	construction rather than as an independent dynamical result.
	A direct comparison with the Model~A value (mean-field $z=2$, defined with
	respect to physical time) would conflate two different clocks.
	
	\subsection{Fokker--Planck equation}
	
	So far we have worked at the deterministic, $N\to\infty$ level.
	To capture finite-size fluctuations, we must retain the stochastic nature
	of the discrete growth process.
	The two leading conditional moments of the one-step increment
	$\Delta\phi = \phi' - \phi$ are computed from the exact update
	Eq.~\eqref{eq:discrete_update}.
	The first moment is the drift:
	\begin{align}
		\E[\Delta\phi\mid\phi]
		&= \frac{\mu(\phi)}{N}+\mathcal{O}(N^{-2}), \label{eq:first_moment}
	\end{align}
	where the $1/N$ prefactor reflects the fact that each admission changes
	$\phi$ by a fraction of order $1/N$.
	
	The second moment requires the conditional variance of $\delta$.
	Since $\delta\in\{0,1\}$ is Bernoulli with success probability
	$Q_+(\phi)$, we have $\E[\delta^2\mid\phi]=\E[\delta\mid\phi]=Q_+(\phi)$,
	and therefore
	\begin{align}
		\E[(\Delta\phi)^2\mid\phi]
		&= \frac{\E[(2\delta-1-\phi)^2\mid\phi]}{(N+1)^2} \notag\\
		&= \frac{Q_+(\phi)(1-\phi)^2 + [1-Q_+(\phi)](1+\phi)^2}{N^2}
		\notag\\
		&\quad + \mathcal{O}(N^{-3}) \notag\\
		&\equiv \frac{\sigma^2(\phi)}{N^2}+\mathcal{O}(N^{-3}).
		\label{eq:second_moment}
	\end{align}
	Defining the noise intensity
	\begin{equation}\label{eq:sigma_def_main}
		\sigma^2(\phi) \equiv Q_+(\phi)(1-\phi)^2+[1-Q_+(\phi)](1+\phi)^2,
	\end{equation}
	the second moment is $\mathcal{O}(N^{-2})$, consistent with a diffusion
	process on the slow logarithmic timescale.
	
	The Kramers--Moyal expansion of the master equation for the probability
	density $P(\phi,\tau)$ truncates at second order (the higher moments are
	$\mathcal{O}(N^{-3})$ or smaller), yielding the Fokker--Planck equation
	on logarithmic time:
	\begin{equation}\label{eq:fpe}
		\begin{split}
			\frac{\partial P}{\partial \tau}
			&= -\frac{\partial}{\partial\phi}
			\left[\mu(\phi)P\right]  \\
			&\quad
			+ \frac{1}{2N(\tau)}
			\frac{\partial^2}{\partial\phi^2}
			\left[\sigma^2(\phi)P\right],
		\end{split}
	\end{equation}
	where $N(\tau)=N_0 e^\tau$.
	Equivalently, the stochastic dynamics is described by the It\^{o} Langevin
	equation
	\begin{equation}\label{eq:langevin}
		\diff\phi_\tau
		=
		\mu(\phi_\tau)\,\diff\tau
		+
		\frac{\sigma(\phi_\tau)}{\sqrt{N(\tau)}}\,\diff W_\tau ,
	\end{equation}
	where $W_\tau$ is a standard Wiener process.
	Two features of this Langevin equation are noteworthy.
	First, the noise is multiplicative: $\sigma(\phi)$ depends on the current
	state $\phi$, with the admission variance largest near $\phi=0$ (where
	$Q_+\approx 1/2$).
	We note that $\sigma^2$ does \emph{not} vanish at the boundaries for
	$\eta>0$: $\sigma^2(\pm1)=4[1-Q_+(\pm1)]>0$, because misjudged
	admissions of the minority type remain possible; the boundary noise
	vanishes only in the zero-noise limit $\eta\to0$, a point that becomes
	important for the absorbing-state discussion of Sec.~\ref{sec:dp}.  Second, the noise amplitude decays with group size as
	$1/\sqrt{N}$, confirming that fluctuations are suppressed in the
	thermodynamic limit---the deterministic dynamics of Eq.~\eqref{eq:drift}
	is recovered exactly as $N\to\infty$.
	
	\subsection{Quasipotential and Kramers escape rate}
	\label{sec:quasipotential}
	
	The growth process is nonstationary: $N(\tau)$ increases and the noise
	weakens as the group grows, so the process possesses no stationary
	distribution in the strict sense.
	To characterize fluctuations at large but finite size we therefore use a
	\emph{frozen-$N$ auxiliary diffusion}: we fix $N$ in
	Eq.~\eqref{eq:fpe} and study the stationary solution of the resulting
	time-homogeneous equation, which describes the instantaneous fluctuation
	cost at size $N$.
	For large $N$, the distribution is sharply peaked around the deterministic
	attractors, so we employ the WKB (Wentzel--Kramers--Brillouin) ansatz
	\begin{equation}\label{eq:WKB_ansatz}
		P_{\text{st}}(\phi) \simeq A(\phi)\,e^{-N S(\phi)},
	\end{equation}
	where $S(\phi)$ is the quasipotential and $A(\phi)$ is a slowly varying
	prefactor.  Substituting this ansatz into the stationary Fokker--Planck
	equation and retaining only the leading terms in $1/N$ yields the
	Hamilton--Jacobi equation
	\begin{equation}\label{eq:HJ}
		\mu(\phi)S'(\phi)
		+\frac{\sigma^2(\phi)}{2}\,[S'(\phi)]^2=0.
	\end{equation}
	The trivial solution $S'(\phi)=0$ corresponds to the deterministic
	steady states.  The nontrivial branch,
	$S'(\phi) = -2\mu(\phi)/\sigma^2(\phi)$, integrates to the
	Freidlin--Wentzell quasipotential
	\begin{equation}\label{eq:S_formal}
		S(\phi) = -2\int_{1}^{\phi} \frac{\mu(y)}{\sigma^2(y)}\,\diff y.
	\end{equation}
	The quasipotential controls finite-size behavior:
	local minima of $S$ correspond to metastable macrostates, the global minimum
	selects the thermodynamic state in the $N\to\infty$ limit, and barrier heights
	$\Delta S$ determine exponentially long escape times through factors of the
	form $\exp(N\Delta S)$.
	
	In the language of large deviation theory \cite{Touchette2009},
	$S(\phi)$ plays the role of a rate function for the frozen-$N$ diffusion,
	\begin{equation}\label{eq:LDP}
		\prob(\phi_N \approx \phi) \asymp e^{-N S(\phi)},
	\end{equation}
	and should be interpreted as the finite-$N$ fluctuation cost, not as a
	stationary thermodynamic potential of the growth process itself.
	A rigorous large deviation principle for the original urn-type stochastic
	approximation (e.g.\ via the G\"{a}rtner--Ellis theorem
	\cite{Gartner1977,Dembo2010}) is plausible but beyond the scope of this
	paper.
	
\begin{figure}[htbp]
		\centering
		\includegraphics[width=\columnwidth]{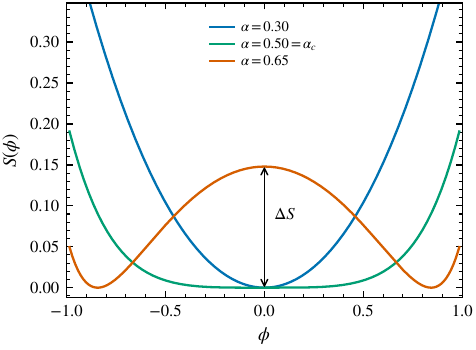}
		\caption{Freidlin--Wentzell quasipotential $S(\phi)$ for $m=2$, $f=1/2$.
			In the ordered phase ($\alpha>\alpha_c$), two symmetric minima
			separated by a barrier $\Delta S$ at $\phi=0$.}
		\label{fig:quasipotential}
	\end{figure}

	For $m=2$, the integrand is rational and $S(\phi)$ evaluates in elementary
	functions (\ref{app:quasipotential_explicit}).
	The result is shown in Fig.~\ref{fig:quasipotential}.
	
	For $f<1/2$, the positive branch $\phi_+$ is metastable.
	For the frozen-$N$ diffusion at size $N$, the mean first-passage time
	follows the Kramers-type formula \cite{Freidlin2012}:
	\begin{equation}\label{eq:kramers}
		\begin{split}
			\tau_{\text{escape}}
			&\sim
			\frac{2\pi}{\sqrt{|\mu'(\phi_+)|\,\mu'(\phi_s)}}  \\
			&\quad\times \exp\!\big(N\,\Delta S\big),
		\end{split}
	\end{equation}
	where $\phi_s$ is the saddle and $\Delta S \equiv S(\phi_s) - S(\phi_+)$
	[for the state-dependent noise of Eq.~\eqref{eq:langevin} the prefactor
	acquires additional factors of $\sigma^2(\phi_+)$ and $\sigma^2(\phi_s)$;
	the exponential dependence is unaffected].
	Near the saddle-node bifurcation,
	$\Delta S \propto (\alpha-\alpha_{\rm sn})^{3/2}$
	\cite{Strogatz2018} (derived from the universal saddle-node normal form in
	\ref{app:rg_scaling}), implying exponentially long metastable
	lifetimes for large groups when $\alpha>\alpha_{\rm sn}(f,m)$.
	In the actual growth process the noise weakens further as $N$ increases, so
	$e^{N\Delta S}$ is the instantaneous escape scale at size $N$ and escapes
	become progressively rarer as the group grows; a quantitative
	first-passage analysis of the time-inhomogeneous process is left open.
	
	\subsection{Logarithmic-time relaxation scaling}
	\label{sec:fss}

	We test the mean-field description through a scaling collapse on
	logarithmic growth time.
	We use this name deliberately: $N$ is simultaneously system size and
	growth time, so the analysis below is finite-\emph{growth-time} scaling
	of the relaxation, not finite-size scaling in the equilibrium or
	absorbing-state sense.
	[The Binder cumulant
	$U_4=1-\langle\phi^4\rangle/3\langle\phi^2\rangle^2$ \cite{Binder1981}
	would provide a standard complementary diagnostic; we do not pursue it
	here.]
	The crucial point---and a direct corollary of the central thesis that
	$\tau=\ln N$ is simultaneously the physical time and the RG scale---is that
	the relevant scaling variable is the elapsed log-time
	$\tau=\ln(N/N_0)$, not $N$ itself.
	Because the founding condition $\phi(N_0)=1$ breaks the symmetry from the
	outset, the ensemble-averaged order parameter follows the deterministic
	relaxation of Eq.~\eqref{eq:drift} on log-time, and finite-size effects are
	controlled by the elapsed log-time $\tau$.
	The appropriate scaling ansatz is therefore
	\begin{equation}\label{eq:fss_form}
		\phi(N,\eta)
		= \tau^{-\beta/\nu}\,
		\mathcal{G}\!\big((\eta-\eta_c)\,\tau\big),
		\qquad \frac{\beta}{\nu}=\frac12,
	\end{equation}
	where $\tau=\ln(N/N_0)$, with $\beta=1/2$ and the correlation-time
	exponent $\nu=1$ both established
	above (Sec.~\ref{sec:dynamic_z}), and $z=1$ so that no separate dynamic
	exponent enters.
	The exponents have a transparent origin.
	At criticality ($\alpha=\alpha_c$) the linear term of the drift vanishes and
	Eq.~\eqref{eq:drift_expansion} reduces to
	$\diff\phi/\diff\tau=-\tfrac13 m\alpha_c^3(m^2-1)\,\phi^3$, whose solution
	decays algebraically in log-time,
	\begin{equation}\label{eq:critical_relaxation}
		\phi(\tau)\simeq
		\left[\tfrac{2}{3}m\alpha_c^3(m^2-1)\,\tau\right]^{-1/2}
		\;\xrightarrow{\;m=2\;}\;
		\sqrt{2}\,\tau^{-1/2},
	\end{equation}
	fixing the amplitude exponent $\beta/\nu=1/2$ and the asymptotic
	universal value $\mathcal{G}(0)\to\sqrt{2}$ ($\tau\to\infty$) for
	$m=2$.
	Away from criticality, the deviation $t=\alpha-\alpha_c$ enters through the
	relaxation rate $\lambda_\tau=m|t|$, so the crossover is governed by the
	dimensionless combination $t\,\tau\propto(\eta-\eta_c)\,\tau$, i.e.\ $\nu=1$.
	This log-time relaxation scaling replaces the naive
	equilibrium-volume estimate ($\phi\sim N^{-1/4}$); the Monte Carlo data
	below clearly favor the logarithmic form.

	We verified these exponents through Monte Carlo simulation of the exact
	discrete-time growth process defined in Sec.~\ref{sec:model}.
	Each simulation run initializes a group with $N_0=10$ founding members
	(all type $+1$) and grows it to a target size $N$ by sequentially
	drawing candidates, evaluating them through the unanimity rule, and
	admitting those who pass.  The key simulation parameters are:
	$m=2$, $f=1/2$, group sizes $N$ ranging from $2\times 10^3$ to
	$2\times 10^5$, and evaluation noise $\eta$ spanning $[0.20, 0.30]$
	in steps of $0.005$---a range that brackets the critical point
	$\eta_c=0.25$.  For each parameter combination $(N,\eta)$, we average
	over $n_{\text{runs}}=200$ independent realizations to obtain the
	ensemble-averaged polarization $\langle\phi\rangle$ and cohesion
	$\langle C\rangle = (1+\langle\phi\rangle)/2$.
	The standard error of the mean is smaller than $1\%$ of the
	cohesion value for the largest system sizes.
	An additional run at the critical point $\eta=\eta_c$ extends to
	$N=2\times 10^{8}$.
	Evaluators are sampled with replacement, matching the analytic treatment;
	realizations use independent pseudorandom streams, and error bars denote
	the standard error of the ensemble mean over the $n_{\text{runs}}$
	realizations.
	Rejected candidates advance neither $N$ nor $\phi$; on the growth clock
	they are equivalently absorbed into the conditional admission law
	$\delta\sim\mathrm{Bernoulli}(Q_+(\phi))$, and we verified that both
	implementations give statistically identical results.
	Simulation code and data are available from the authors upon request.
	We verified that doubling the founding size to $N_0=20$ shifts the
	rescaled data by at most a few percent, within the slow
	$\mathcal{O}(1/\tau)$ corrections.

\begin{figure*}[t]
		\centering
		\includegraphics[width=\textwidth]{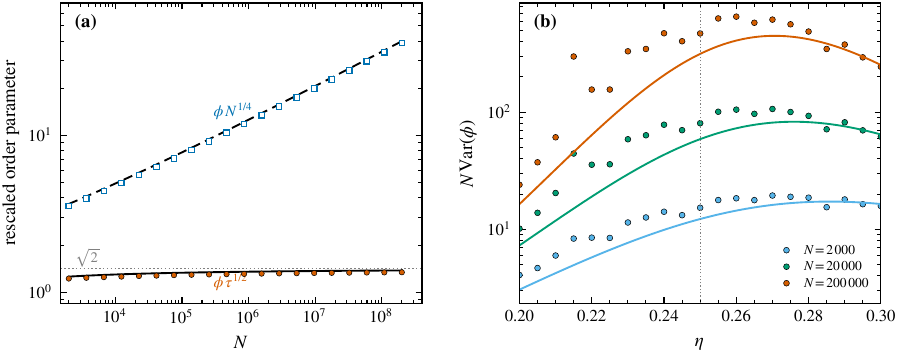}
		\caption{Stochastic-sector tests ($m=2$, $f=1/2$; $200$ realizations
			in~(a), $10^{3}$ in~(b)).
			(a)~Critical-point decay up to $N=2\times10^{8}$:
			$\phi\,\tau^{1/2}$ (circles) remains bounded and approaches
			$\sqrt{2}$ (dotted), whereas the naive volume rescaling
			$\phi N^{1/4}$ (open squares) grows without bound; solid and
			dashed lines are the parameter-free deterministic predictions.
			(b)~Ensemble variance of $\phi$ versus $\eta$ for three group
			sizes (symbols), scaled by $N$, compared with the parameter-free
			linearized prediction of Eq.~\eqref{eq:variance_ode} (lines).
			The prediction captures magnitude and shape; the residual excess
			reflects nonlinear corrections beyond the linearization.}
		\label{fig:crit_fluct}
	\end{figure*}

\begin{figure*}[t]
		\centering
		\includegraphics[width=\textwidth]{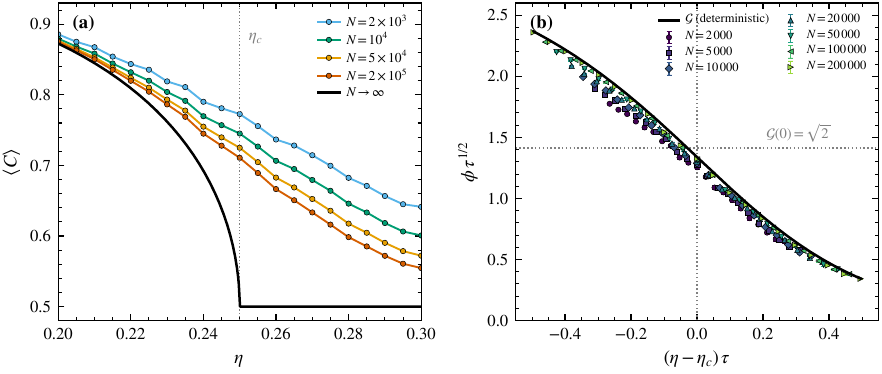}
		\caption{Finite-size scaling analysis of Monte Carlo simulations ($m=2$,
			$f=1/2$, $2\times 10^3 \le N \le 2\times 10^5$, $200$ realizations
			per point).
			(a)~Ensemble-averaged cohesion $\langle C\rangle$ versus evaluation
			noise $\eta$ for various group sizes, compared with the analytical
			$N\to\infty$ curve (black solid line).
			(b)~Log-time scaling collapse with $\beta/\nu=1/2$ and $z=1$:
			$\phi\,\tau^{1/2}$ versus $(\eta-\eta_c)\,\tau$, with
			$\tau=\ln(N/N_0)$ the elapsed log-time.
			Data for all sizes collapse onto the parameter-free deterministic
			master curve $\mathcal{G}$ (solid black, obtained by integrating
			the drift from $\phi=1$); the dotted line marks the asymptotic
			critical value $\mathcal{G}(0)=\sqrt{2}$, approached as
			$\tau\to\infty$, consistent with the mean-field fixed-point
			exponents on logarithmic time.
			Error bars are smaller than the markers.}
		\label{fig:fss_collapse}
	\end{figure*}

	Figure~\ref{fig:fss_collapse}(a) displays the raw finite-size crossover:
	as $N$ increases, the ensemble-averaged cohesion $\langle C\rangle$
	sharpens toward the analytical $N\to\infty$ curve (solid black line),
	which follows from solving $\arctanh(2C-1)=2\arctanh(\alpha(2C-1))$.
	Near the critical point, the curves for different $N$ fan out---the
	signature of a diverging correlation time---while far from criticality
	they converge rapidly.
	Panel (b) demonstrates the scaling collapse: data for all system sizes,
	spanning two decades from $N=2\times 10^3$ to $2\times 10^5$, collapse onto
	a single master curve when plotted as
	$\phi\,\tau^{\beta/\nu}$ versus $(\eta-\eta_c)\,\tau$, with
	$\beta/\nu=1/2$ and $\tau=\ln(N/N_0)$.
	The collapsed data fall precisely on the parameter-free deterministic
	scaling function $\mathcal{G}$ obtained by integrating the drift
	Eq.~\eqref{eq:drift} from $\phi=1$; at the critical point the curve
	approaches the asymptotic value $\mathcal{G}(0)=\sqrt{2}$ from below as
	$\tau\to\infty$ [Eq.~\eqref{eq:critical_relaxation}], the residual
	offset at the simulated sizes being the slow $\mathcal{O}(1/\tau)$
	correction visible in the inset.
	For comparison, the same collapse attempted with naive equilibrium-volume
	exponents ($\phi N^{1/4}$ vs $(\eta-\eta_c)N^{1/2}$) fails: at criticality
	the rescaled order parameter grows as $N^{1/4}$ instead of remaining
	bounded, because the critical decay is logarithmic,
	$\phi\sim\tau^{-1/2}$, not a power of $N$.
	Figure~\ref{fig:crit_fluct}(a) makes the comparison direct, using a
	dedicated critical-point run extending over five decades to
	$N=2\times10^{8}$: $\phi\,\tau^{1/2}$ remains essentially constant,
	while $\phi N^{1/4}$ grows by more than an order of magnitude.

	The collapse of the ensemble \emph{mean} probes the deterministic sector;
	to probe the stochastic sector directly we compare the measured ensemble
	variance of $\phi$ with the linearized small-noise (van Kampen-type)
	prediction obtained by integrating
	\begin{equation}\label{eq:variance_ode}
		\frac{\diff\,\mathrm{Var}(\phi)}{\diff\tau}
		= 2\mu'(\bar\phi)\,\mathrm{Var}(\phi)
		+ \frac{\sigma^2(\bar\phi)}{N_0}\,e^{-\tau}
	\end{equation}
	along the deterministic trajectory $\bar\phi(\tau)$, with
	$\mathrm{Var}=0$ at $\tau=0$ [the factor $e^{-\tau}/N_0=1/N(\tau)$ is
	the instantaneous noise strength of Eq.~\eqref{eq:langevin}].
	Figure~\ref{fig:crit_fluct}(b) shows that this parameter-free prediction
	reproduces the magnitude and the $\eta$ dependence of the measured
	variance over four orders of magnitude in $N\,\mathrm{Var}(\phi)$, with
	median excess factors of $1.2$--$1.7$ growing with $\tau$; the largest
	deviations (up to a factor $\approx 4$ at the biggest size) occur in a
	narrow window just below $\eta_c$, where the onset of bistability makes
	the linearization marginal.
	The fluctuations are thus quantitatively tied to the same microscopic
	noise amplitude $\sigma^2(\phi)$ that enters the quasipotential,
	independently of the deterministic collapse.

	Because the master curve is itself the integrated deterministic flow, the
	collapse should be read as a stringent test of the mean-field description
	and of the logarithmic scaling variable---not as an independent
	measurement of fluctuation exponents: what is verified is that the
	ensemble-averaged dynamics follows the deterministic mean-field flow with
	$\beta=1/2$, $\nu=1$, and $z=1$ on logarithmic time.
	We also note that $\nu$ is a correlation-\emph{time} exponent; the
	all-to-all model has no spatial correlation length.

	\section{Microscopic origin of criticality: correlated verdicts}
	\label{sec:clt}

	Let evaluator $k$ return a binary verdict $X_k\in\{0,1\}$ with
	$\E[X_k] = p_+(\phi)$.
	Under unanimity, the admission probability is
	$\mathcal{A}_m(\phi) = \E[\prod_{k=1}^{m} X_k]$.
	For conditionally independent evaluators,
	$\mathcal{A}_m(\phi) = f[p_+(\phi)]^m + (1-f)[p_-(\phi)]^m$.

	The verdicts are correlated through the common candidate type:
	$\Cov(X_1, X_2) \neq 0$.
	Conditioned on the candidate type $c_j$, the evaluators' verdicts are
	independent Bernoulli variables with success probability
	$p_+(\phi)$ if $c_j=+1$ and $p_-(\phi)$ if $c_j=-1$.
	Unconditionally, they are coupled by the shared hidden variable $c_j$.
	Thus the unanimity rule produces a \emph{mixture of product measures}:
	\begin{equation}\label{eq:mixture_measure}
		\begin{split}
		\E[X_1 X_2 \cdots X_m]
		&= f \cdot \E[X_1\cdots X_m \mid c_j=+1] \\
		&\quad + (1-f) \cdot \E[X_1\cdots X_m \mid c_j=-1].
		\end{split}
	\end{equation}
	Conditioned on $c_j$, the evaluators are independent, so each conditional
	expectation factorizes:
	\begin{equation}\label{eq:product_moment}
		\mathcal{A}_m(\phi)
		=
		\E\!\left[\prod_{k=1}^{m}X_k\right]
		=
		f[p_+(\phi)]^m+(1-f)[p_-(\phi)]^m.
	\end{equation}

	The structure of this expression is the key to understanding why $m\ge 2$
	is required for criticality.  For $m=1$, the admission probability
	$\mathcal{A}_1(\phi) = f p_+(\phi) + (1-f)p_-(\phi)
	= \tfrac12[1+(2f-1)\alpha\phi]$ is \emph{linear} in
	$\phi$ and, at $f=1/2$, independent of $\phi$ altogether.  For $m\ge 2$, the powers
	$[p_\pm(\phi)]^m$ enhance the linear response of $Q_+$ and generate
	higher-order terms in $\phi$; as shown below, the enhanced linear
	coefficient provides the positive feedback necessary for a symmetry-breaking
	instability, while the higher-order terms saturate it.

	To make this quantitative, define the \emph{excess admission probability}
	\begin{equation}\label{eq:excess_def}
		\Delta_m(\phi) \equiv Q_+(\phi) - p_+(\phi),
	\end{equation}
	which measures how much the $m$-evaluator consensus rule deviates from
	the single-evaluator baseline $p_+(\phi)$.
	For $m=1$, $Q_+(\phi)=p_+(\phi)$ by definition, so $\Delta_1(\phi)=0$
	identically.
	For $m\ge 2$ and $f=1/2$, expanding $Q_+(\phi)$ for small $\phi$:
	\begin{align}
		Q_+(\phi)
		&= \frac{[1+\alpha\phi]^m}
		{[1+\alpha\phi]^m+[1-\alpha\phi]^m} \notag\\
		&= \frac{1}{2}
		+ \frac{m\alpha}{2}\phi
		- \frac{m\alpha^3(m^2-1)}{6}\phi^3
		+ \mathcal{O}(\phi^5).
		\label{eq:Q_expansion}
	\end{align}
	Because $Q_+(-\phi)=1-Q_+(\phi)$ at $f=1/2$, the function
	$Q_+(\phi)-\tfrac12$ is \emph{odd} in $\phi$: only odd powers appear, and
	there is no $\phi^2$ term.
	Since $p_+(\phi)=\tfrac12(1+\alpha\phi)=\tfrac12+\tfrac{\alpha}{2}\phi$,
	the difference at leading order is purely \emph{linear},
	\begin{equation}\label{eq:excess_expansion}
		\Delta_m(\phi)
		= \frac{(m-1)\alpha}{2}\,\phi
		- \frac{m\alpha^3(m^2-1)}{6}\phi^3
		+ \mathcal{O}(\phi^5).
	\end{equation}

	This is the microscopic origin of criticality.
	The excess \emph{vanishes identically} for $m=1$ (where $Q_+=p_+$),
	explaining the absence of a finite critical noise (the $m=1$ anomaly of
	Sec.~\ref{sec:beyond}).
	For $m\ge 2$, the excess carries a positive \emph{linear} coefficient
	$(m-1)\alpha/2$: the consensus rule systematically amplifies the current
	majority.  Adding the feedback $2\Delta_m(\phi)\simeq (m-1)\alpha\phi$ to
	the single-evaluator baseline drift $2p_+(\phi)-1-\phi=(\alpha-1)\phi$
	reproduces the net linear coefficient $m\alpha-1$ of
	Eq.~\eqref{eq:mu_prime_0}; the disordered state $\phi=0$ is destabilized
	precisely when this amplification overcomes the restoring force
	$-2\eta\phi=(\alpha-1)\phi$, i.e.\ for $\alpha>1/m$.
	The factor $m-1$ is the combinatorial remnant of collective evaluation: a
	single evaluator merely samples the composition and cannot amplify it,
	whereas two or more evaluators are correlated through the shared candidate
	type and generate the additional alignment $\propto(m-1)\alpha\phi$.
	In the language of Sec.~\ref{sec:bayes}, $\Delta_m$ is the evidence
	pooled through the shared candidate: conditioning on unanimity converts $m$
	noisy readings of the composition into a single amplified one.
	The leading \emph{nonlinear} term, which saturates the growth and bounds the
	ordered branch, appears only at cubic order with coefficient
	$\propto m\alpha^3(m^2-1)$ [Eq.~\eqref{eq:drift_expansion}].
	This analysis also reveals why the critical reliability $\alpha_c=1/m$
	decreases with $m$: more evaluators extract more signal from the noisy
	verdicts, so a lower reliability (higher noise) suffices to sustain order.

	\section{P\'{o}lya urn duality and martingale convergence}
	\label{sec:polya}

	\subsection{Mapping}

	For $m=1$, the replacement matrix is
	$\mathbf{R} = \bigl(\begin{smallmatrix}1-\eta & \eta\\ \eta & 1-\eta\end{smallmatrix}\bigr)$
	(Friedman urn).
	For $m\ge 2$, the effective replacement rule becomes nonlinear in the
	composition, corresponding to a consensus-reinforced generalized
	P\'{o}lya urn: on the growth clock, each added ``ball'' is of type $+1$
	with the composition-dependent probability $Q_+(\phi)$ of
	Eq.~\eqref{eq:Q_plus}.
	This nonlinear reinforcement is the urn-theoretic form of consensus
	amplification.

	\subsection{Theorem}

	Let $\mathcal{F}_N$ be the natural filtration and
	$g(C) = Q_+(2C-1) - C$.
	The following statement is a standard application of
	stochastic-approximation theory \cite{Pemantle2007}, recorded here for
	completeness.

	\emph{Theorem 1} (Convergence).
	For $m\ge 2$ and $f=1/2$, $C_N$ converges almost surely as $N\to\infty$.
	For $\alpha \le 1/m$ the limit is $C^*=1/2$.
	For $\alpha > 1/m$ the limit lies almost surely in the two-point set
	$\{C^-,C^+\}$ of stable zeros of $g$, where
	$C^{\pm}=\tfrac12[1\pm\phi^*]$ and $\phi^*$ is the unique positive
	solution of Eq.~\eqref{eq:arctanh_eigen}; the founding condition
	$C_{N_0}=1$ favors the cohesive branch, with
	$\prob(\lim_N C_N=C^+)\to 1$ as $N_0\to\infty$.

	\textit{Proof sketch.}
	The composition obeys the exact recursion
	\begin{equation}\label{eq:robbins_monro}
		C_{N+1}=C_N+\frac{g(C_N)}{N+1}+\zeta_{N+1},
		\qquad
		\E[\zeta_{N+1}\mid\mathcal{F}_N]=0,
	\end{equation}
	which is a Robbins--Monro stochastic-approximation scheme with step size
	$1/(N+1)$ and martingale-difference noise $\zeta_{N+1}$.  Because a single
	admission changes $C_N$ by at most $1/(N+1)$, the noise is bounded by
	$|\zeta_{N+1}|\le 2/(N+1)$, so the squared increments are summable,
	$\sum_{N}\E[\zeta_{N+1}^2\mid\mathcal{F}_N]\le\sum_N 4/(N+1)^2<\infty$.
	The two standard Robbins--Monro conditions, $\sum_N 1/(N+1)=\infty$ and
	$\sum_N \E[\zeta_{N+1}^2]<\infty$, are therefore met, and the
	Robbins--Siegmund / Kushner--Clark theorem guarantees that $C_N$ converges
	almost surely to the stable-equilibrium set of the limiting ODE
	$\dot C=g(C)$.  Because early fluctuations of order $1/N_0$ carry the
	trajectory across $\phi=0$ with strictly positive probability, the
	founding condition $C_{N_0}=1$ selects the cohesive branch only
	probabilistically: $\prob(\lim_N C_N=C^+)\to 1$ as $N_0\to\infty$,
	while convergence to the stable set $\{C^-,C^+\}$ (for $\alpha>1/m$) and
	to $1/2$ (for $\alpha\le 1/m$) is almost sure.
	(Full proof, including $L^2$-martingale convergence, in
	\ref{app:martingale}.)
	This provides a rigorous bridge between the stochastic growth process and
	the deterministic fixed-point analysis.

	For $m=1$, $g(C) = -\eta(2C-1)$, so $C^* = 1/2$ for any $\eta>0$.

	\section{A renormalization-group analogy}
	\label{sec:rg}

	The deterministic growth flow admits a suggestive reading in
	renormalization-group (RG) language, with the group size $N$ playing the
	role of the flow scale.
	We develop this analogy here and state precisely where it ends: no
	fluctuation degrees of freedom are integrated out, and the couplings
	$(m,\alpha)$ do not flow---what flows is the order parameter itself.

	In conventional RG, one integrates out short-wavelength degrees of freedom
	to obtain an effective description at longer scales.
	In the growth process, each admission step adds one member to the group,
	incrementally ``coarse-graining'' the collective opinion: the new member's
	type is determined by sampling the existing composition through the noisy
	evaluation process, and the updated composition $\phi(N+1)$ is a nonlinear
	function of $\phi(N)$.
	In this restricted sense, the growth process enacts the coarse-graining
	itself.

	Formally, the RG transformation $\mathcal{R}_b$ coarse-graining from group
	size $N$ to group size $N/b$ (i.e., integrating out a factor $b$ of the
	membership) is obtained by integrating the deterministic flow
	$\diff\phi/\diff\ell = \mu(\phi)$ over the scale interval
	$\ell \equiv \ln b$:
	\begin{equation}\label{eq:RG_transform}
		\begin{split}
			\mathcal{R}_b[\phi]
			&= \phi + \mu(\phi)\ln b  \\
			&\quad
			+ \tfrac12\mu(\phi)\mu'(\phi)(\ln b)^2
			+ \mathcal{O}((\ln b)^3).
		\end{split}
	\end{equation}
	Equation~\eqref{eq:RG_transform} is an exact property of the deterministic
	flow in $\tau=\ln N$; we emphasize, however, that it transports the order
	parameter at fixed couplings, and is not a Wilsonian renormalization of
	$(m,\alpha)$.

	The analog of a beta function is the deterministic drift:
	\begin{equation}\label{eq:beta_RG}
		\beta(\phi) \equiv \frac{\diff\phi}{\diff\ell} = \mu(\phi),
	\end{equation}
	with $\ell \equiv \ln b$ the logarithmic RG scale.
	The fixed points of the RG flow are precisely the fixed points of the
	deterministic dynamics:
	\begin{itemize}[nosep]
		\item \textbf{Gaussian fixed point}: $\phi^*=0$, stable for
		$\alpha<1/m$ (disordered phase).  This is the infrared-stable
		fixed point when noise dominates.
		\item \textbf{Critical fixed point}: $\phi^*=0$ at $\alpha=1/m$,
		marginal.  The linearized flow $\beta(\phi)\approx
		(m\alpha-1)\phi$ vanishes, and the leading nonlinear terms
		determine the universal behavior.
		\item \textbf{Ordered fixed points}: $\phi^*=\pm\phi^*(\alpha)$,
		stable for $\alpha>1/m$.  These are the infrared-stable fixed
		points in the ordered phase.
	\end{itemize}

	Linearizing around the Gaussian fixed point gives the flow rate
	$\beta'(\phi^*=0)=m\alpha-1=m(\alpha-\alpha_c)$.
	The relevant scaling field is the deviation $t\equiv\alpha-\alpha_c$, and
	the correlation length defined through $\lambda_\tau^{-1}\sim|t|^{-\nu}$
	diverges with $\nu=1$ (Sec.~\ref{sec:dynamic_z}).
	Equivalently, the relevant RG eigenvalue is $y_t=1/\nu=1$, so that under a
	rescaling by a factor $b$ the field transforms as $t\to b^{y_t}t$ with
	$y_t=1$, consistent with \ref{app:rg_scaling}.

	Two caveats delimit the content of this identification.
	First, the exponent $\nu=1$ follows from the linear-stability eigenvalue
	of the ODE at the Gaussian fixed point; the RG language repackages this
	result rather than deriving it independently.
	Second, $z=1$ expresses the fact that the flow scale and the physical
	time are the same variable $\tau=\ln N$, so scale invariance and dynamic
	scaling coincide by construction (Sec.~\ref{sec:dynamic_z}).
	What the analogy does capture is that the coarse-graining is enacted by
	the dynamics itself: each admission step updates the composition according
	to the flow generated by $\beta(\phi)=\mu(\phi)$, with no external
	transformation applied.
	A genuine Wilsonian calculation---integrating the $\mathcal{O}(1/N)$
	fluctuations of Eq.~\eqref{eq:langevin} to obtain a one-loop flow of the
	Landau couplings---is an interesting open problem.

	For the scaling of the quasipotential barrier near the continuous
	transition, $\Delta S \sim \varepsilon^{2}$ with
	$\varepsilon=\alpha-\alpha_c$, see \ref{app:rg_scaling}.


	\section{Irreversibility: a Kullback--Leibler measure of time asymmetry}
	\label{sec:stoch_thermo}

	The growth process is inherently irreversible: members are admitted but
	never removed, and each admission decision consumes information about the
	candidate and the group composition.
	To quantify this irreversibility we adapt the trajectory formalism of
	stochastic thermodynamics \cite{Seifert2012}, with an important caveat
	spelled out below.

	Consider a single admission step in which a candidate of type
	$c_j\in\{\pm1\}$ is admitted as member $\delta\in\{0,1\}$.
	The forward probability of this event, conditioned on the current state
	$\phi$, is $\mathcal{P}(\delta\mid\phi) = Q_+(\phi)^\delta
	[1-Q_+(\phi)]^{1-\delta}$ (on the growth clock, where each step is an
	admission).
	A trajectory entropy production is conventionally defined as the log-ratio
	of forward to backward path probabilities.
	The growth process, however, has no removal mechanism and hence no bona
	fide time-reversed dynamics; any ``backward'' process must be chosen as a
	reference.
	Here we compare the realized admission with the outcome-flipped one---a
	convention rather than a true time reversal, so the quantity below is a
	\emph{reference-dependent measure of dynamical time asymmetry} rather than
	a unique thermodynamic entropy production:
	\begin{equation}\label{eq:traj_entropy}
		\begin{split}
		\Delta s_N(\delta)
		&= \ln\frac{\mathcal{P}(\delta\mid\phi)}
		{\mathcal{P}(1-\delta\mid\phi')} \\
		&= \delta\ln\frac{Q_+(\phi)}{1-Q_+(\phi')}
		+ (1-\delta)\ln\frac{1-Q_+(\phi)}{Q_+(\phi')}.
		\end{split}
	\end{equation}
	For large $N$, $\phi'\approx\phi$, and averaging over the Bernoulli
	distribution of $\delta$ gives the ensemble-averaged entropy production
	rate per admission:
	\begin{equation}\label{eq:entropy_rate}
		\begin{split}
			\langle\dot{s}\rangle
			&= Q_+(\phi)\ln\frac{Q_+(\phi)}{1-Q_+(\phi)}
			+ [1-Q_+(\phi)]\ln\frac{1-Q_+(\phi)}{Q_+(\phi)} \\[4pt]
			&= (2Q_+-1)\,\ln\frac{Q_+}{1-Q_+}.
		\end{split}
	\end{equation}
	Equation~\eqref{eq:entropy_rate} is the Kullback--Leibler divergence rate
	between the admission distribution $\mathrm{Bernoulli}(Q_+)$ and its
	outcome-flipped counterpart $\mathrm{Bernoulli}(1-Q_+)$.
	In the ordered phase ($\phi\neq 0$), $Q_+\neq 1/2$ and
	$\langle\dot{s}\rangle > 0$: the growth process has a systematic
	directionality that favors the majority type.
	At the critical point, $Q_+(\phi)\to 1/2$ as $\phi\to 0$, and the entropy
	production rate vanishes as $\phi^2$.
	In the steady state, $\langle\dot{s}\rangle\propto\phi^{*2}\propto
	(\alpha-\alpha_c)$ just above the transition and vanishes below it, so
	$\langle\dot{s}\rangle$ has a kink at $\alpha_c$ and its
	$\alpha$-derivative a finite jump---formally analogous to the mean-field
	specific-heat discontinuity, and providing a purely dynamical signature of
	the transition.

	A thermodynamic-uncertainty-type tradeoff suggests itself for the
	cumulative cohesion $J_N = \sum_{k=N_0}^{N} C_k$: in its standard form,
	the thermodynamic uncertainty relation (TUR)
	\cite{Barato2015,Gingrich2016} states
	\begin{equation}\label{eq:TUR}
		\frac{\Var(J_N)}{\langle J_N\rangle^2}
		\cdot \langle\Delta s_{\text{total}}\rangle
		\ge 2k_B.
	\end{equation}
	We caution that the standard proofs of Eq.~\eqref{eq:TUR} assume a
	stationary time-homogeneous Markov process, whereas the growth process is
	nonstationary and $\langle\Delta s_{\text{total}}\rangle$ is reference
	dependent; whether a TUR holds here is open.
	Heuristically, near criticality the variance of $J_N$ is enhanced while
	the accumulated time asymmetry grows slowly
	($\langle\dot{s}\rangle\propto\phi^2$), so any TUR-type bound would be
	far from saturation: critical growth produces large fluctuations at small
	irreversibility cost.

	\section{Information geometry}
	\label{sec:info_geo}

	The steady-state distribution $P_{\text{st}}(\phi; \alpha, f) \propto
	e^{-N S(\phi; \alpha, f)}$ defines a statistical manifold.
	The Fisher information \cite{Amari2000}:
	\begin{equation}\label{eq:fisher}
		g_{\mu\nu} = \int_{-1}^{1} \partial_\mu\ln P_{\text{st}}\,
		\partial_\nu\ln P_{\text{st}}\,
		P_{\text{st}}(\phi)\,\diff\phi.
	\end{equation}

	We evaluate the metric in the large-$N$ Laplace approximation, taking care to
	distinguish the two control parameters: they play the roles of
	temperature and field, respectively, and behave differently near the
	transition.
	In a single well the distribution is asymptotically Gaussian,
	$\phi\sim\mathcal{N}\!\big(\phi^*,\,[N S''(\phi^*)]^{-1}\big)$, and the Fisher
	information of a Gaussian family with parameter-dependent mean and variance is
	$g_{\theta\theta}=N S''(\phi^*)\,(\partial_\theta\phi^*)^2
	+\tfrac12(\partial_\theta\ln S'')^2$, whose leading term in $N$ is
	\begin{equation}\label{eq:fisher_general}
		g_{\theta\theta}\simeq N\,S''(\phi^*)\,
		\left(\frac{\partial\phi^*}{\partial\theta}\right)^{\!2}.
	\end{equation}

	\emph{Reliability (temperature-like) direction.}  Near the continuous
	transition $S(\phi)$ has the Landau form
	$S\simeq (r/\sigma_0^2)\phi^2+\cdots$
	(\ref{app:quasipotential_explicit}), so at the ordered minimum
	$S''(\phi^*)\sim|r|\sim|\alpha-\alpha_c|$, while
	$\phi^*\propto(\alpha-\alpha_c)^{1/2}$ gives
	$(\partial_\alpha\phi^*)^2\sim|\alpha-\alpha_c|^{-1}$.  The two factors
	\emph{cancel}, leaving
	\begin{equation}\label{eq:fisher_alpha}
		g_{\alpha\alpha}\simeq N\,|\alpha-\alpha_c|^{-1}\cdot|\alpha-\alpha_c|
		= \mathcal{O}(N),
	\end{equation}
	a finite amplitude rather than a divergence---the information-geometric
	counterpart of the mean-field result that the specific heat has only a jump
	($\alpha_{\rm sp}=0$) in the reliability (temperature) channel.

	\emph{Bias (field-like) direction.}  The genuinely divergent component is the
	one conjugate to the symmetry-breaking field $h=\tfrac12\ln[f/(1-f)]$.  A small
	$h$ shifts the disordered-phase peak to $\phi^*=\chi h$ with
	$\chi=1/r=1/(1-m\alpha)$, so $\partial_h\phi^*=\chi$ and
	$S''(0)=2r/\sigma_0^2$, giving
	\begin{equation}\label{eq:fisher_h}
		g_{hh}\simeq N\,S''(0)\,\chi^2
		= \frac{2N}{\sigma_0^2}\,\chi
		\;\propto\; N\,|\alpha-\alpha_c|^{-1}
		\xrightarrow{\alpha\to\alpha_c}\infty.
	\end{equation}
	Thus it is the field-conjugate Fisher information, not the reliability one,
	that diverges with the susceptibility exponent $\gamma=1$, in exact parallel
	with equilibrium criticality (where $g_{HH}\propto\chi$ diverges while
	$g_{TT}\propto C_V$ does not).
	This divergence is the information-theoretic counterpart of susceptibility:
	arbitrarily small changes in the candidate bias become statistically
	distinguishable near the transition, the finite-size cutoff being set by the
	critical window $|\alpha-\alpha_c|\sim(\ln N)^{-1}$ of Sec.~\ref{sec:fss}.
	A full Ruppeiner scalar-curvature analysis \cite{Ruppeiner1995} requires
	the mixed component $g_{\alpha h}$ and is left for future work.

	\section{Outlook: spatial generalizations and absorbing states}
	\label{sec:dp}
	
	All results presented so far assume \emph{all-to-all} evaluator selection:
	each evaluator is chosen uniformly at random from the entire group,
	independently of spatial or network proximity.
	When evaluators are instead chosen from a spatial neighborhood---for
	instance, on a regular lattice or a spatially embedded network---the
	dynamics may acquire a qualitatively new ingredient: an \emph{absorbing
	state}.
	This section is an outlook; as we explain, the absorbing-state scenario is
	a conjecture about modified or limiting versions of the model, not a
	property of the model at its critical point.

	Consider a spatial domain in which all members of one type (say, $-1$) have
	been eliminated.
	For $\eta>0$ such a region is \emph{not} absorbing: by
	Eq.~\eqref{eq:micro}, each $+1$ evaluator approves a $-1$ candidate with
	probability $\eta$, so a pure $+1$ region admits a $-1$ candidate with
	probability $\eta^m>0$ per arrival.
	A true absorbing state exists only in the zero-noise limit $\eta\to 0$
	(equivalently $\alpha\to 1$), or in modified local rules in which the
	misjudgment probability vanishes with the local minority fraction.
	Since the mean-field critical point sits at $\eta_c=1/2-1/(2m)>0$, far
	from the zero-noise limit, the considerations below concern such limiting
	or modified models.
	We further note that for $f=1/2$ at $\eta=0$ the two fully polarized
	states $\phi=\pm1$ are \emph{simultaneously} absorbing, so the symmetric
	variant may fall in a two-absorbing-state (voter/parity-conserving) class
	rather than ordinary directed percolation, depending on how the limit is
	taken.

	In a zero-noise spatial variant, the dynamics would satisfy the
	Janssen--Grassberger conditions for directed-percolation (DP) universality
	\cite{Janssen1981,Grassberger1982,Hinrichsen2000}: a scalar order
	parameter $\phi(\mathbf{r},t)$, short-range spreading under local
	evaluator selection, and (at $\eta=0$) an absorbing fully polarized state
	with no additional conservation law.
	
	The continuum field theory for the spatial variant is
	\begin{equation}\label{eq:field_theory}
		\partial_t \phi = D\nabla^2\phi + \mu(\phi)
		+ \sqrt{\sigma^2(\phi)}\,\xi(\mathbf{r},t),
	\end{equation}
	where $D$ is an effective diffusion constant set by the range of evaluator
	selection, $\mu(\phi)$ is the arctanh-derived drift of Eq.~\eqref{eq:drift},
	and $\xi$ is spatiotemporal white noise.
	An important caveat concerns the noise structure: the amplitude inherited
	from the mean-field process, $\sigma^2(\phi)$ of
	Eq.~\eqref{eq:sigma_def_main}, does \emph{not} vanish at $\phi=1$
	[$\sigma^2(1)=4\,(1-Q_+(1))>0$ for $\eta>0$], whereas absorbing-state
	field theories require multiplicative noise that vanishes in the absorbing
	state ($\propto\sqrt{\rho}$ in Reggeon field theory \cite{Cardy1996}).
	Equation~\eqref{eq:field_theory} as written therefore does \emph{not}
	belong to the DP class; the correct noise structure must be rederived for
	the zero-noise local model.
	
	If a modified spatial variant does realize a true absorbing state, its
	critical exponents should be those of the DP universality class.
	In two spatial dimensions, the established DP values are
	$\beta\approx 0.583$, $z\approx 1.76$, and $\nu_\perp\approx 0.733$
	\cite{Hinrichsen2000}---all distinct from the mean-field values
	$\beta=1/2$, $z=1$, $\nu=1$ derived in this paper.
	The all-to-all model corresponds to the $D\to\infty$ (or equivalently,
	infinite-dimensional) limit, where mean-field theory becomes exact and
	the absorbing-state physics is suppressed by the global connectivity.
	
	The crossover from mean field to DP is a natural direction for future
	work.  As the spatial range of evaluator selection is reduced, the
	exponents may flow from the mean-field values reported here toward
	absorbing-state universality classes appropriate to the spatial
	dimension---provided the absorbing state is realized as discussed above.  A systematic
	$\varepsilon$-expansion around the upper critical dimension
	$d_c=4$ of DP, using the arctanh nonlinearity as the microscopic input,
	together with numerical simulations of one- and two-dimensional variants,
	would settle the universality class; we leave both as open problems.
	
	\section{Discussion and outlook}
	\label{sec:discussion}
	
	\subsection{Summary}
	
	We have developed a systematic theoretical framework for a Hamiltonian-free
	growth-driven phase transition, organized in three layers.
	
	The \textbf{core theory} (Secs.~\ref{sec:model}--\ref{sec:dynamics})
	centers on the arctanh fixed-point equation
	$\arctanh(\phi^*) = m \cdot \arctanh(\alpha\phi^*) + h$,
	which compactly encodes the entire fixed-point structure and from which
	all measurable quantities follow.
	Its physical content is a log-odds balance: the group's composition
	reproduces the Bayesian posterior of its own noisy evaluations, and order
	persists when the inference gain $m\alpha$ of the consensus rule exceeds
	unity (Sec.~\ref{sec:bayes}).
	The explicit Landau coefficients $r=1-m\alpha$, $u=m\alpha^3(m^2-1)/3$,
	and $v=-m\alpha^5(2m^4-5m^2+3)/15$ are fully microscopic.
	The log-time relaxation exponent $z=1$ reflects the identification of
	the logarithmic time $\tau=\ln N$ with the flow scale.
	The frozen-$N$ Freidlin--Wentzell quasipotential and Kramers-type estimates
	characterize instantaneous fluctuation costs and metastability.
	Monte Carlo simulations collapse onto the parameter-free deterministic
	master curve (Fig.~\ref{fig:fss_collapse}), consistent with the mean-field
	exponents $\beta=1/2$, $\nu=1$, and $z=1$ on logarithmic time.
	
	The \textbf{mathematical foundations} (Secs.~\ref{sec:clt}--\ref{sec:rg})
	place the phase transition on rigorous footing in three complementary
	ways.  The microscopic analysis shows that the excess
	admission probability $\Delta_m(\phi)\simeq\tfrac12(m-1)\alpha\phi$ is the
	microscopic mechanism generating the linear positive feedback, and
	that $m=1$ annihilates it exactly.  The P\'{o}lya-urn martingale
	construction proves almost-sure convergence to the deterministic fixed
	points, establishing that the growth process is a stochastic approximation
	to the ODE $\dot C=g(C)$.  The RG analogy shows that the deterministic
	flow can be read as a coarse-graining of the composition with drift
	$\beta(\phi)=\mu(\phi)$, with the caveat that the couplings themselves
	do not flow.
	
	The \textbf{complementary perspectives} (Secs.~\ref{sec:stoch_thermo}--\ref{sec:dp})
	extend the framework in three directions.
	The irreversibility analysis shows that the time-asymmetry rate vanishes
	as $\phi^2$ and exhibits a kink at the critical point---the analog of the
	mean-field specific-heat discontinuity---while the applicability of
	thermodynamic uncertainty relations to this nonstationary process remains
	an open question.
	Information geometry reveals that the field-conjugate Fisher information
	diverges as $N|\alpha-\alpha_c|^{-1}$, making arbitrarily small candidate
	biases statistically distinguishable at criticality.
	The spatial generalization is left as a conjecture: a true absorbing state
	requires the zero-noise limit or modified local rules, and the resulting
	universality class (DP, voter-like, or other) remains open.
	
	A structural comparison reveals a nuanced relationship with the equilibrium
	paradigm.  While both the growing-group model and the mean-field Ising model
	share the same static universality class, their mathematical architectures
	differ fundamentally: the nested arctanh structure
	$\arctanh(\phi)\propto\arctanh(\alpha\phi)$ has no Ising counterpart.
	The model furthermore exhibits features without direct equilibrium
	counterpart: the $m=1$ anomaly (no finite critical noise for single
	evaluators, arising from the vanishing of the excess admission probability
	rather than from low dimensionality), and, for $f<1/2$, a permanent memory
	of the founding condition---the irreversible growth provides no channel to
	relax out of the metastable branch, so the discontinuous jump is dictated
	by the initial condition rather than by an equilibrium coexistence
	construction.  The discontinuity line terminates at an ordinary mean-field
	critical point at $f=1/2$ with a cusp (spinodal) structure---and, notably,
	\emph{not} a tricritical point, since the quartic Landau coupling
	$u(\alpha_c)=(m^2-1)/3m^2$ remains strictly positive for the unanimity
	rule.
	
	Table~\ref{tab:paradigm} provides the complete paradigm comparison.
	
	\begin{table}[htbp]
		\caption{Paradigm comparison: equilibrium vs.\ nonequilibrium growth.}
		\label{tab:paradigm}
		\scriptsize
		\setlength{\tabcolsep}{2pt}
		\renewcommand{\arraystretch}{1.22}
		\begin{tabular}{@{}L{0.46\columnwidth}L{0.47\columnwidth}@{}}
			\toprule
			Equilibrium (Ising) & Growth (this work) \\
			\midrule
			Hamiltonian $H(\{\sigma_i\})$ & Admission rule $Q_+(\phi)$ \\
			Partition function $Z = \Tr e^{-\beta H}$ & Transition probability $Q_+(\phi)$ \\
			Free energy $F = -k_BT\ln Z$ & $\mathcal{F}_{\eff} = -\int\mu(\phi)\diff\phi$ \\
			Boltzmann factor $e^{-\beta H}/Z$ & Quasipotential $e^{-N S(\phi)}$ \\
			Temperature $T$ & Evaluation noise $\eta = (1-\alpha)/2$ \\
			External field $H$ & Candidate bias $h = \frac12\ln[f/(1-f)]$ \\
			Coordination number $q$ & Evaluator count $m$ \\
			Thermodynamic limit $N\to\infty$ & RG limit $\tau = \ln N\to\infty$ \\
			Landau theory of $T_c$ & Bifurcation theory of $\alpha_c=1/m$ \\
			Fluctuation--dissipation theorem & Fokker--Planck equation \\
			Model A dynamics & RG-like growth flow \\
			Boltzmann entropy & Trajectory irreversibility measure \\
			Specific heat singularity & Fisher information divergence \\
			\bottomrule
		\end{tabular}
	\end{table}
	
	\subsection{On the meaning of ``Hamiltonian-free''}
	
	A conceptual clarification is in order.
	This paper constructs an effective potential $\mathcal{F}_{\eff}$, a
	quasipotential $S(\phi)$, and an exact RG equation from the nonequilibrium
	growth dynamics.
	One may ask whether this amounts to introducing a Hamiltonian through
	the back door.
	
	The answer is no, and the distinction can be made precise at three levels.
	At the \emph{microscopic} level, the Hamiltonian $H(\{\sigma_i\})$ in
	equilibrium statistical mechanics is \emph{prescribed}: it specifies the
	interaction energies between all pairs of spins, and the Boltzmann weight
	$e^{-\beta H}$ is the fundamental postulate from which all equilibrium
	properties follow.
	In the growth model, the corresponding object is the admission rule
	$Q_+(\phi)$, which is not an energy function but a \emph{conditional
		probability} derived from the microscopic evaluation process.
	No interaction energy between group members is ever postulated---only the
	probability that a candidate passes the evaluation, which follows from
	the independent-noisy-verdict model of Eq.~\eqref{eq:micro}.
	
	At the \emph{macroscopic} level, $\mathcal{F}_{\eff}(\phi) =
	-\int_1^\phi \mu(y)\,\diff y$ is \emph{not} a free energy in the
	thermodynamic sense: it is a Lyapunov function of the deterministic drift,
	constructed a posteriori from the dynamics.  In equilibrium, the free energy
	$F = -k_B T \ln Z$ is the \emph{generating function} from which all
	observables are derived; here, $\mathcal{F}_{\eff}$ is a \emph{summary} of
	the deterministic flow and has no associated partition function.  One cannot
	compute, say, the specific heat from $\mathcal{F}_{\eff}$ by taking
	temperature derivatives, because there is no temperature---only the
	evaluation noise $\eta$, which enters $\mathcal{F}_{\eff}$ through the
	drift $\mu(\phi)$ in a fundamentally different way than temperature enters
	the Boltzmann factor.
	
	At the \emph{fluctuation} level, the quasipotential $S(\phi)$ is not a
	Boltzmann factor but a large-deviation \emph{rate function}.
	It satisfies $S(\phi) = \lim_{N\to\infty} -N^{-1}\ln P_{\text{st}}(\phi)$,
	encoding the exponential cost of fluctuations away from the deterministic
	attractor, rather than the energetic cost of a microstate.
	The crucial conceptual difference is that $S(\phi)$ is \emph{derived} from
	the stochastic dynamics, whereas the Boltzmann factor $e^{-\beta H}/Z$ is
	\emph{postulated} as the starting point.
	
	That the resulting structure---Landau theory, critical
	exponents, scaling relations, RG flow---so closely parallels equilibrium
	theory is not a weakness but the central finding: it demonstrates that
	the mathematical architecture of critical phenomena is largely independent
	of the equilibrium postulate.  A purely nonequilibrium, Hamiltonian-free
	growth process can generate the same universality classes, the same types
	of phase transitions, and the same scaling structure as an equilibrium
	system, because these are consequences of symmetry, dimensionality, and
	the analytic structure of the fixed-point equations, not of the existence
	of a Hamiltonian.
	
	\subsection{Extensions}
	
	Several directions are open for future work.
	\emph{Majority voting} (threshold $k$ out of $m$) introduces a new
	parameter $k/m$.
	\emph{Member departure} creates a birth--death process with nontrivial
	finite-$N$ stationarity.
	\emph{Field-theoretic formulation} via Doi--Peliti \cite{Doi1976,Peliti1985}
	or response-functional methods \cite{Janssen1976} would enable systematic
	$\varepsilon$-expansion.
	\emph{Experimental tests} of the predicted critical group size, metastable
	lifetime, and the critical-isotherm law $|\phi^*|\propto|f-1/2|^{1/3}$
	along the $\alpha=\alpha_c$ line are feasible in controlled settings with
	human participants
	or synthetic agent-based simulations.
	
	\subsection{Conclusion}
	
	We have demonstrated that the full machinery of critical phenomena---continuous
	and discontinuous phase transitions, critical exponents, scaling relations,
	universality classes, history-dependent metastability, spinodal (cusp)
	structure, quasipotentials, and convergence theorems---can be constructed
	without a Hamiltonian.
	The central insight is the arctanh representation
	$\arctanh(\phi^*) = m\cdot\arctanh(\alpha\phi^*) + h$, which encodes the
	entire fixed-point structure of the consensus-driven growth process in a
	single compact equation---and which we identified as the exact log-odds
	balance of a self-consistent inference loop, with the prior bias as the
	external field and the collective gain $m\alpha$ as the control
	parameter.
	Around this core, the three-layer theoretical framework provides a complete
	picture: the core theory (Landau coefficients, Ising comparison, beyond-Ising
	phenomena, log-time relaxation scaling) establishes \emph{what} the model
	predicts;
	the mathematical foundations (correlated verdicts, P\'{o}lya urn, the RG
	analogy) explain \emph{why} the transition occurs at the microscopic level
	and in what sense the growth process can be read as a flow; and the
	complementary perspectives (irreversibility, information geometry, and the
	conjectured absorbing-state generalizations) connect the model to broader
	themes in nonequilibrium physics.
	Together, these results establish that Hamiltonian-free growth can serve as
	an organizing principle for critical phenomena in a broad class of complex
	systems driven by sequential accretion, evaluation, and consensus.
	
	\section*{CRediT authorship contribution statement}
	Xingfu Ke: Visualization, Validation, Methodology, Conceptualization.  Fanyuan Meng: Writing – review \& editing, Writing – original draft, Supervision, Methodology, Conceptualization.
	\section*{Declaration about generative AI}
	A large language model was used for linguistic refinement during manuscript preparation. All contents were carefully reviewed and edited by the authors, who bear full responsibility for the published work.
	\section*{Declaration of competing interest}
	The authors declare that they have no known competing financial interests or personal relationships that could have appeared to influence the work reported in this paper.
	\section*{Acknowledgments}
This work is supported by the National Natural Science Foundation of China (Grant Nos. 12505044 and 52374013).

	\appendix
	
	\section{Noise intensity \texorpdfstring{$\sigma^2(\phi)$}{sigma squared of phi}}
	\label{app:noise}
	
	The paper uses group size $N$ as the clock.  Thus each elementary step is an
	accepted candidate, and $\delta=1$ with probability $Q_+(\phi)$.
	The exact jump is
	\begin{equation}
		\Delta\phi
		=
		\frac{2\delta-1-\phi}{N+1}.
	\end{equation}
	The first Kramers--Moyal coefficient is therefore
	\begin{align}
		\E[\Delta\phi\mid\phi]
		&=
		\frac{2Q_+(\phi)-1-\phi}{N+1}
		=
		\frac{\mu(\phi)}{N}
		+\mathcal{O}(N^{-2}).
	\end{align}
	Let
	\begin{equation}\label{eq:sigma_growth_clock}
		\sigma^2(\phi)
		=
		Q_+(\phi)(1-\phi)^2
		+[1-Q_+(\phi)](1+\phi)^2 .
	\end{equation}
	The raw second moment is then
	\begin{equation}
		\E[(\Delta\phi)^2\mid\phi]
		=
		\frac{\sigma^2(\phi)}{(N+1)^2}
		=
		\frac{\sigma^2(\phi)}{N^2}
		+\mathcal{O}(N^{-3}).
	\end{equation}
	Equivalently,
	\begin{equation}
		\sigma^2(\phi)
		=
		1+\phi^2-2\phi[2Q_+(\phi)-1].
	\end{equation}
	
	For the symmetric one-evaluator case, $m=1$ and $f=1/2$,
	\begin{equation}
		\begin{aligned}
			Q_+(\phi)
			&=
			\frac{1+\alpha\phi}{2},\\
			\sigma^2(\phi)
			&=
			1+\phi^2-2\alpha\phi^2
			=
			1+(1-2\alpha)\phi^2 .
		\end{aligned}
	\end{equation}
	
	\section{Explicit quasipotential for \texorpdfstring{$m=2$}{m=2}}
	\label{app:quasipotential_explicit}
	
	For $m=2$, $f=1/2$, the integrand $\mu(\phi)/\sigma^2(\phi)$ is a rational
	function.
	First,
	\begin{equation}
		Q_+(\phi)
		=
		\frac{(1+\alpha\phi)^2}
		{(1+\alpha\phi)^2+(1-\alpha\phi)^2}
		=
		\frac{1+2\alpha\phi+\alpha^2\phi^2}
		{2(1+\alpha^2\phi^2)} .
	\end{equation}
	Hence the deterministic drift reduces to
	\begin{equation}\label{eq:app_mu_m2}
		\mu(\phi)
		=
		2Q_+(\phi)-1-\phi
		=
		\frac{\phi(2\alpha-1-\alpha^2\phi^2)}
		{1+\alpha^2\phi^2}.
	\end{equation}
	Using Eq.~\eqref{eq:sigma_growth_clock}, the noise intensity is
	\begin{equation}\label{eq:app_sigma_m2}
		\sigma^2(\phi)
		=
		\frac{
			1+(1+\alpha^2-4\alpha)\phi^2+\alpha^2\phi^4
		}
		{1+\alpha^2\phi^2}.
	\end{equation}
	Therefore
	\begin{equation}\label{eq:app_ratio_m2}
		\frac{\mu(\phi)}{\sigma^2(\phi)}
		=
		\frac{\phi(2\alpha-1-\alpha^2\phi^2)}
		{1+(1+\alpha^2-4\alpha)\phi^2+\alpha^2\phi^4}.
	\end{equation}
	The quasipotential
	\begin{equation}
		S(\phi)
		=
		-2\int_1^\phi
		\frac{y(2\alpha-1-\alpha^2y^2)}
		{1+(1+\alpha^2-4\alpha)y^2+\alpha^2y^4}
		\,\diff y
	\end{equation}
	is therefore elementary: the substitution $u=y^2$ converts it to an integral
	of a rational function.  In the general symmetric case,
	\begin{align}
		\mu(\phi)
		&=
		(m\alpha-1)\phi
		-\frac{m\alpha^3(m^2-1)}{3}\phi^3
		+\mathcal{O}(\phi^5),\\
		\sigma^2(\phi)
		&=
		\sigma_0^2+\sigma_2^2\phi^2+\mathcal{O}(\phi^4),
	\end{align}
	where $\sigma_0^2=1$ in the growth clock.  Thus near the continuous
	transition the quasipotential has the Landau-like form
	\begin{equation}
		S(\phi)
		=
		\frac{1-m\alpha}{\sigma_0^2}\phi^2
		+\frac{m\alpha^3(m^2-1)}{6\sigma_0^2}\phi^4
		+\cdots ,
	\end{equation}
	which explains why the minima of $S$ coincide with the deterministic stable
	fixed points.
	
	\section{Landau coefficients and special limits}
	\label{app:landau_details}
	
	The identity
	\begin{equation}
		2Q_+(\phi)-1
		=
		\tanh\!\left[m\,\arctanh(\alpha\phi)\right]
	\end{equation}
	gives the symmetric drift
	$\mu(\phi)=\tanh[m\,\arctanh(\alpha\phi)]-\phi$.
	With $\theta=\arctanh(\alpha\phi)$,
	\begin{align}
		\theta
		&=
		\alpha\phi+\frac{\alpha^3\phi^3}{3}
		+\frac{\alpha^5\phi^5}{5}
		+\mathcal{O}(\phi^7),\\
		\tanh(m\theta)
		&=
		m\theta-\frac{m^3\theta^3}{3}
		+\frac{2m^5\theta^5}{15}
		+\mathcal{O}(\theta^7).
	\end{align}
	Collecting powers gives
	\begin{align}
		\mu(\phi)
		&=
		(m\alpha-1)\phi
		-\frac{m\alpha^3(m^2-1)}{3}\phi^3 \notag\\
		&\quad
		+\frac{m\alpha^5(2m^4-5m^2+3)}{15}\phi^5
		+\mathcal{O}(\phi^7).
	\end{align}
	Since $\mathcal{F}'_{\eff}(\phi)=-\mu(\phi)$, this yields
	\begin{align}
		r &= 1-m\alpha,\\
		u &= \frac{m\alpha^3(m^2-1)}{3},\\
		v &= -\frac{m\alpha^5(2m^4-5m^2+3)}{15}.
	\end{align}
	For $m=1$, all nonlinear coefficients vanish and
	\begin{equation}
		\frac{\diff\phi}{\diff\tau}
		=
		(\alpha-1)\phi
		=
		-2\eta\phi .
	\end{equation}
	Because $\tau=\ln(N/N_0)$, the exact solution is
	\begin{equation}
		\phi(N)=\phi(N_0)\left(\frac{N_0}{N}\right)^{2\eta}.
	\end{equation}
	This is the precise sense in which the $m=1$ transition is pushed to
	$\eta_c=0$: for every $\eta>0$, the order parameter decays algebraically.
	
	\section{Saddle-node line for \texorpdfstring{$f<1/2$}{f<1/2}}
	\label{app:saddle_node_details}
	
	For $f\neq1/2$, write
	$h=\frac12\ln[f/(1-f)]$ and
	\begin{equation}
		\Phi(\phi;\alpha,m)
		=
		\arctanh\phi
		-m\,\arctanh(\alpha\phi).
	\end{equation}
	Fixed points satisfy $\Phi(\phi;\alpha,m)=h$.
	A saddle-node occurs when the horizontal line $h$ is tangent to $\Phi$, so
	\begin{equation}
		\Phi(\phi_c;\alpha_{\rm sn},m)=h,
		\qquad
		\partial_\phi\Phi(\phi_c;\alpha_{\rm sn},m)=0 .
	\end{equation}
	The tangency condition gives
	\begin{equation}
		\frac{1}{1-\phi_c^2}
		=
		\frac{m\alpha_{\rm sn}}{1-\alpha_{\rm sn}^2\phi_c^2},
	\end{equation}
	or
	\begin{equation}
		\phi_c^2
		=
		\frac{m\alpha_{\rm sn}-1}{\alpha_{\rm sn}(m-\alpha_{\rm sn})}.
	\end{equation}
	Substituting the positive root into $\Phi=h$ gives the metastability boundary
	in parametric form:
	\begin{equation}
		f_c(\alpha,m)
		=
		\frac{1}
		{1+\exp[-2\Phi(\phi_c;\alpha,m)]}.
	\end{equation}
	
	\section{Martingale convergence: complete proof}
	\label{app:martingale}
	
	Let $\mathcal{F}_N = \sigma(\{C_k\}_{k=N_0}^N)$ and
	$g(C) = Q_+(2C-1) - C$.
	Define the compensated process
	$M_N = C_N - \sum_{k=N_0}^{N-1} g(C_k)/(k+1)$.
	Since $\E[C_{N+1}\mid\mathcal{F}_N]=C_N+g(C_N)/(N+1)$, we have
	$\E[M_{N+1}\mid\mathcal{F}_N]=M_N$, so $M_N$ is an
	$\mathcal{F}_N$-martingale whose increments
	$\Delta M_N=M_{N+1}-M_N=\zeta_{N+1}$ are bounded, $|\Delta M_N|\le 2/(N+1)$.

	\emph{Step 1 ($L^2$ convergence).}  Because the increments are orthogonal,
	\begin{equation}\label{eq:martingale_L2}
		\begin{split}
		\E[M_N^2]-\E[M_{N_0}^2]
		&=\sum_{k=N_0}^{N-1}\E[(\Delta M_k)^2] \\
		&\le\sum_{k=N_0}^{\infty}\frac{4}{(k+1)^2}<\infty .
		\end{split}
	\end{equation}
	The squared increments are \emph{summable} (a constant, independent of $N$),
	so $\sup_N\E[M_N^2]<\infty$: $M_N$ is an $L^2$-bounded martingale.  By the
	Doob martingale convergence theorem it converges almost surely and in
	$L^2$ to a \emph{finite} limit $M_\infty$ (note: to a finite random variable,
	not necessarily to zero).  A na\"{i}ve Azuma--Hoeffding bound of the form
	$\exp(-\varepsilon^2 N/8)$ does \emph{not} apply here precisely because the
	sum of squared increments converges rather than growing linearly in $N$.

	\emph{Step 2 (stochastic approximation).}  Convergence of $M_N$ implies that
	the accumulated noise $\sum_k\zeta_{k+1}$ converges, so the exact recursion
	\eqref{eq:robbins_monro} is a Robbins--Monro scheme with decreasing step
	$1/(N+1)$ satisfying $\sum_N 1/(N+1)=\infty$ and
	$\sum_N\E[\zeta_{N+1}^2]<\infty$.  By the Robbins--Siegmund and
	Kushner--Clark theorems \cite{Pemantle2007}, $C_N$ then converges almost
	surely to a stable equilibrium of the mean ODE $\dot C=g(C)$; the
	convergence of urn-type reinforcement recursions to stable zeros is the
	classical result of Athreya--Karlin \cite{Athreya1968}.

	\emph{Step 3 (stability selection).}  The stable zeros are determined by the
	sign of $g(C)$ between zeros.  For $f=1/2$ and $\alpha>1/m$, $C=1/2$ is
	unstable ($g'(1/2)=\tfrac12(m\alpha-1)>0$) while the symmetric zeros
	$C^{\pm}$ are stable; nonconvergence to the unstable point is ruled out by
	Pemantle-type theorems \cite{Pemantle2007}, so the limit lies in
	$\{C^-,C^+\}$ almost surely.  Because a finite run of early $-1$
	admissions occurs with strictly positive probability, the event
	$\{\lim_N C_N = C^-\}$ cannot be excluded for finite $N_0$; standard
	stochastic-approximation estimates give
	$\prob(\lim_N C_N=C^+)\to 1$ as $N_0\to\infty$.
	For $\alpha\le1/m$,
	$C=1/2$ is the unique stable zero, proving convergence to the random-group
	state.
	
	\section{RG scaling of the quasipotential barrier}
	\label{app:rg_scaling}
	
	Because the model has no spatial structure ($d_{\eff}=0$), the barrier
	carries no volume prefactor, and its dependence on
	$\varepsilon=\alpha-\alpha_c$ follows directly from the Landau form of the
	quasipotential (\ref{app:quasipotential_explicit}).
	Writing $S(\phi)\simeq\tilde r\,\phi^2+\tilde u\,\phi^4$ with
	$\tilde r=(1-m\alpha)/\sigma_0^2=-m\varepsilon/\sigma_0^2$ and
	$\tilde u=m\alpha^3(m^2-1)/(6\sigma_0^2)$, the ordered minima sit at
	$\phi^{*2}=-\tilde r/(2\tilde u)$ and the barrier separating them from
	$\phi=0$ is
	\begin{equation}
		\Delta S=\frac{\tilde r^{\,2}}{4\tilde u}
		\;\propto\;\varepsilon^{2},
	\end{equation}
	the quadratic power being the mean-field analog of the singular
	free-energy scaling $|t|^{2-\alpha_{\rm sp}}$ with specific-heat exponent
	$\alpha_{\rm sp}=0$.
	
	\subsection{Saddle-node normal form and the $3/2$ barrier law}
	
	At a first-order collapse ($f<1/2$), the dynamics near the saddle-node
	bifurcation reduces to the universal normal form
	\begin{equation}\label{eq:normal_form}
		\dot x = a - x^2 + \sqrt{D/N}\,\xi(t),
	\end{equation}
	where $x$ is the coordinate along the center manifold, $a \propto
	\alpha - \alpha_{\rm sn}(f,m)$ measures the distance from the bifurcation
	point,
	$D$ is the effective noise intensity, and $\xi(t)$ is white noise.
	
	The corresponding Freidlin--Wentzell quasipotential is obtained by
	integrating the deterministic part:
	\begin{equation}\label{eq:normal_form_S}
		S(x) = -\frac{2}{D}\int_{x_0}^{x} (a - y^2)\,\diff y
		= \frac{2}{D}\!\left(\frac{x^3}{3} - a x\right) + \const.
	\end{equation}
	For $a > 0$, the deterministic flow $\dot x = a-x^2$ has a stable fixed
	point at $x_+ = +\sqrt{a}$ and an unstable saddle at $x_- = -\sqrt{a}$;
	correspondingly $S(x)$ has its minimum at $x_+$ and its maximum at $x_-$.
	The barrier height for escape from the metastable state over the saddle is
	\begin{equation}\label{eq:barrier_3over2}
		\begin{split}
			\Delta S = S(x_-) - S(x_+)
			&= \frac{2}{D}\!\left(\frac{2a^{3/2}}{3}\right)
			- \frac{2}{D}\!\left(-\frac{2a^{3/2}}{3}\right) \\[4pt]
			&= \frac{8}{3D}\,a^{3/2}.
		\end{split}
	\end{equation}
	
	Expressing the bifurcation parameter in terms of the physical control
	variables, $a \propto \alpha - \alpha_{\rm sn}(f,m)$ [the metastable
	cohesive branch exists for $\alpha>\alpha_{\rm sn}$ and is annihilated at
	$\alpha=\alpha_{\rm sn}$], we obtain
	$\Delta S \propto (\alpha - \alpha_{\rm sn})^{3/2}$, as quoted in the
	main text.
	Inserting this into the Kramers formula
	\eqref{eq:kramers} yields the scaling of the metastable lifetime:
	\begin{equation}\label{eq:lifetime_scaling}
		\tau_{\text{escape}} \sim
		\exp\!\big(c N (\alpha - \alpha_{\rm sn})^{3/2}\big),
	\end{equation}
	with $c$ a nonuniversal constant.
	The $3/2$ exponent is universal for all saddle-node bifurcations in the
	presence of weak noise \cite{Dykman1994,Berglund2007} and distinguishes the discontinuous
	transition from the continuous one, where $\Delta S \sim \varepsilon^{2}$.
	
	\bibliographystyle{elsarticle-num}

\end{document}